\documentclass[a4paper,11pt]{article}
\usepackage{jheppub} 
\usepackage{orcidlink}

\usepackage{subfigure}
\usepackage{mathtools}
\usepackage{amsmath}
\usepackage{amsfonts}
\usepackage{amssymb}
\usepackage{bbold}
\usepackage{bm}
\usepackage{ esint }
\usepackage{orcidlink}

\newcommand{\bD}{{\bf D}}

\newcommand{\bA}{{\bf A}}

\newcommand{\bP}{{\bf P}}

\newcommand{\br}{{\bf r}}

\newcommand{\bx}{{\bf x}}
\newcommand{\bn}{{\bf n}}

\newcommand{\bk}{{\bf k}}
\newcommand{\bK}{{\bf K}}
\newcommand{\bv}{{\bf v}}
\newcommand{\bp}{{\bf p}}

\newcommand{\bG}{{\bf G}}

\newcommand{\sOm}{{\sf\Omega}}

\newcommand{\bpartial}{\boldsymbol{\partial}}
\newcommand{\bNabla}{\boldsymbol{\nabla}}
\newcommand{\uR}{u_{\rm R}}
\newcommand{\uI}{u_{\rm I}}

\newcommand{\exclude}[1]{{}}
\long\def\exclude#1{}

\newcommand{\GF}{G_{\rm F}}

\title{Theory of neutrino fast flavor evolution.\\
Part I. Linear response theory and stability conditions.}

\author[a]{Damiano F.\ G.\ Fiorillo \orcidlink{0000-0003-4927-9850}} 
\affiliation[a]{Niels Bohr International Academy, Niels Bohr Institute,
University of Copenhagen\\ 2100 Copenhagen, Denmark}

\author[b]{and Georg G.\ Raffelt
\orcidlink{0000-0002-0199-9560}}
\affiliation[b]{Max-Planck-Institut f\"ur Physik (Werner-Heisenberg-Institut)\\ Boltzmannstr.~8, 85748 Garching, Germany}

\abstract{Neutrino-neutrino refraction leads to collective flavor evolution that can include fast flavor conversion, an ingredient still missing in numerical simulations of core-collapse supernovae. We provide a theoretical framework for the linear regime of this phenomenon using the language of response theory. In analogy to electromagnetic waves, we introduce a flavor susceptibility as the linear response to an external flavor field. By requiring self-consistency, this approach leads to the usual dispersion relation for growing modes, but differs from the traditional treatment in that it predicts Landau damping of subluminal collective modes. The new dispersion relation has definite analyticity properties and can be expanded for small growth rates. This approach simplifies and intuitively explains Morinaga's proof of sufficiency for the occurrence of growing modes. We show that weakly growing modes arise as soon as an angular crossing is formed, due to their resonant interaction with individual neutrino modes. For longitudinal plasma waves, a similar resonance causes Landau damping or conversely, the two-stream instability.
}

\begin{document}
\maketitle
\flushbottom

\clearpage

\section{Introduction}

Over the past decades, neutrino flavor conversion has developed from a wild hypothesis~\cite{Gribov:1968kq} to a main-stream topic entering the precision era \cite{Esteban:2020cvm, Capozzi:2021fjo, deSalas:2020pgw}. Incommensurate with this amazing progress, the effect is usually not included in studies of core-collapse supernovae or neutron-star mergers and concomitant nucleosynthesis, even though these events are largely shaped by neutrinos and their flavor-dependent interactions. The main complication derives from neutrino-neutrino refraction that leads to collective forms of flavor conversion \cite{Pantaleone:1992eq, Samuel:1993uw, Samuel:1995ri, Duan:2009cd, Duan:2010bg, Mirizzi:2015eza, Tamborra:2020cul, Capozzi:2022slf, Richers:2022zug,Volpe:2023met}. The corresponding nonlinear kinetic equations of motion, often referred to as quantum kinetic equations (QKEs), are well known \cite{Dolgov:1980cq, Rudsky, Sigl:1993ctk, Sirera:1998ia, Yamada:2000za, Vlasenko:2013fja, Volpe:2013uxl, Serreau:2014cfa, Kartavtsev:2015eva, Fiorillo:2024fnl, Fiorillo:2024wej}, but because of the large separation between flavor evolution scales and overall hydrodynamical ones, a brute-force numerical solution is out of the question. Even ordinary flavor evolution driven by neutrino masses and matter refraction (MSW effect~\cite{Mikheyev:1985zog,Wolfenstein:1977ue}) was usually only studied by post-processing models produced without these effects. Therefore, the community is searching for a physics-informed effective description that would take advantage of the separation of scales and lead to a practical implementation; see, e.g., Refs.~\cite{Bhattacharyya:2020jpj,Zaizen:2022cik,Ehring:2023lcd,Ehring:2023abs,Nagakura:2023jfi,Xiong:2023vcm,Cornelius:2023eop,Johns:2024dbe,Abbar:2024ynh,Fiorillo:2024qbl,Xiong:2024pue}.

Much of the recent focus is on fast flavor conversion, an intriguing effect that does not require neutrino masses or mixing: flavor evolution is driven by the neutrino medium itself in that flavor coherence can spawn yet more flavor evolution, feeding back on itself. This effect probably happens on such short length scales that it can be considered local relative to the much larger hydrodynamical scales in a supernova core or neutron-star merger. The key idea is that classical instabilities (run-away modes of the mean field of flavor coherence) will grow nonlinear and their fragmentation into higher $\bk$-modes can lead to some sort of equilibrium under the constraint of lepton-number conservation. This picture involves large deviations from equilibrium that would then suddenly relax.

These fast flavor instabilities share strong similarities with collective behavior in many other many-body systems, the prime example being longitudinal plasma waves (as we pointed out earlier~\cite{Fiorillo:2023mze} and recap in Appendix~\ref{sec:Landau} for the non-expert), but also collective oscillations in a Fermi liquid.  All of these examples share a fundamental common feature: an ensemble of particles streaming in many directions, described by a ``Vlasov-like'' equation, which feel a collective field generated by all the particles of the ensemble. To see this in our flavor case, we introduce the field of flavor coherence that we call $\psi_{E,\bv}(t,\br)$, which is the flavor off-diagonal element in the usual flavor density matrices $\varrho_{E,\bv}(t,\br)$, where the diagonal elements are the usual occupation numbers. More specifically, in the fast flavor context, the density matrix is for flavor-lepton number (neutrinos minus antineutrinos) that are characterized by energy $E$ and velocity $\bv$, the latter a directional unit vector if we assume the ultrarelativistic limit of vanishing neutrino masses. Moreover, in the massless limit, $E$ drops out from the QKE and after a suitable phase-space integration, we consider the complex function $\psi_{\bv}(t,\br)$, that obeys
\begin{equation}\label{eq:neutrino_eom}
    \big(\partial_t+\bv\cdot\bpartial_\br\big)\psi_{\bv}=
    -i\sqrt{2}\GF\int d^2\bv'\big(1-\bv\cdot\bv'\big)\big(G_\bv\psi_{\bv'}-G_{\bv'}\psi_{\bv}\big).
\end{equation}
$G_\bv$ is the angular distribution of the original undisturbed flavor-lepton flux, taken to be stationary and homogeneous, whereas a dependence on space and time is assumed for~$\psi_\bv$. As usual, one can interpret these quantities geometrically in terms of polarization vectors that represent the trace-free part of the density matrices by $\varrho_\bv=\frac{1}{2}\bigl({\rm Tr}\varrho_\bv+\vec{P}_{\bv}\cdot\vec{\sigma}\bigr)$. We use $\psi_\bv=P^x_\bv+iP^y_\bv$ and the spectrum $G_\bv=P^z_\bv\big|_{t=0}$. The QKEs correspond to the polarization vectors $\vec{P}_\bv$ precessing around a time-varying collective polarization vector generated by all the others in the ensemble.

The analogy with other collective behaviors can now be clarified. The QKEs have a typical Vlasov structure, where the left-hand side (lhs) contains a streaming term describing particles with different velocities, and the right-hand side (rhs) contains the interaction of each neutrino with a collective field generated by all the others. These features are shared by plasma waves, where the collective field is the electric field, and by collective oscillations in Fermi fluids, where the collective field is the refractive potential generated by the short-ranged quasi-particle interactions. The physics behind the instabilities is similar in all these cases. It is only to be expected that the large amount of results collected in decades on these analog systems can yield new and fruitful insights. 

Yet, it is historically curious that the literature on fast flavor conversion has evolved in some sense oppositely to its analogs. Both in plasmas and in Fermi fluids, the initial focus was on \textit{stable} configurations, in which the interaction only manifests itself in the form of collective oscillations such as Langmuir waves. Only later did the attention turn to the more complex possibility of unstable configurations, in which these oscillations could grow exponentially, and even then the accent has always been on configurations that are close to stability. It was implicitly assumed that strongly unstable configurations would likely never show up in Nature. In the case of fast flavor conversions, the approach has been to start from the much more complicated case of unstable systems, considered to be of greater practical interest. 

Standard numerical simulations of core collapse or neutron-star mergers treat neutrino transport without including any form of flavor conversion. Post-processing the output allows one to identify regions of fast flavor instability and thus to judge if a given model is self-consistent. The earliest studies suggested that spherically symmetric models were in this sense self-consistent, but later it turned out that fast flavor instabilities tend to be generic, especially in multi-D simulations, causing the wide-spread interest in this subject and the race for a practical implementation. However, as in plasma physics, a strongly unstable configuration likely never forms so that current discussions may be on some level logically circular. This conundrum was the starting point for Johns' idea that the neutrino flavor field would be in local thermal equilibrium when described by coarse-grained density matrices \cite{Johns:2023jjt}. Recently, he has more explicitly accentuated the traditional approach as logically inconsistent \cite{Johns:2024dbe}, at the same time asserting that the thermodynamic perspective \cite{Johns:2023jjt} was the obvious alternative. On the other hand, based on quasi-linear theory in a toy model, we recently showed explicitly that the system does relax to a spatially-averaged stable configuration, which however needs not be thermodynamical, and provided a practical description of the relaxation mechanism, finding that the system evolves along the edge of stability~\cite{Fiorillo:2024qbl}. 

Whatever the final outcome of these debates, there is a powerful motivation for considering the interacting neutrino system without strong instabilities, or rather, in the linear regime that may be more relevant than previously appreciated. Thus, it is instructive to turn back to the simplest problem of fast flavor evolution near a stable configuration, and take on the lessons of the analog problems in plasma and many-body physics. One can then build up a theory of fast flavor evolution encompassing the stable and unstable cases. This is the subject of this work. 

Our approach revolves around two physical ideas, independent of each other, providing new intuition on the whole subject. First, the stability of the system must be connected with conservation laws. It is well known that a system is mechanically stable when it is in an extremum point of its energy, simply because no other state is accessible without violating the conservation law. A similar point can be made of other conserved quantities, for example angular momentum. Thus, the appearance of instabilities must signal that the system is not in an extremum of some conserved quantity. This provides an intuitive explanation for the role of angular crossings, namely regions where the neutrino lepton number changes sign, in the formation of unstable states. The total polarization vector $\vec{P}_0(t)=\int d^3\br\int d^2\bv \vec{P}_\bv(t,\br)$, analogous to the total angular momentum of our system of precessing spins, is conserved. In the absence of an angular crossing, it is in a maximum state and thus protects the system from instability. An angular crossing implies that states with larger angular momentum exist, allowing for an instability, a necessity argument first advanced in this simple form by Johns \cite{Johns:2024bob},\footnote{However, in his abstract, Johns claims that ergodicity implies that angular crossings are necessary for instability, while in reality ergodicity is not needed at all for the proof of necessity.}
obviating a much more involved dynamical proof by Morinaga \cite{Morinaga:2021vmc} and Dasgupta \cite{Dasgupta:2021gfs}. In Appendix~\ref{app:lepton_number_conservation}  we show the explicit extension of the symmetry argument to slow and collisional instabilities.

If an initial condition is not protected by energy or angular momentum conservation so that other states can be reached without violating these symmetries, suggests that the system is unstable and the necessary condition of a crossing might also suffice. This perspective was taken by Johns based on the speculation of ergodicity \cite{Johns:2024bob}, meaning that the system will equally visit any region of phase space that is not forbidden by conservation laws. It is, however, an assumption, not a proven fact, that the system would have this property. At the very least, this requires that the system has no other symmetries, perhaps invisible to the naked eye. A homogeneous systems has large numbers of Gaudin invariants that make it technically integrable~\cite{Pehlivan:2011hp, Johns:2019izj, Fiorillo:2023mze, Fiorillo:2023hlk}. How do we know if some unobvious combination of such invariants survives the transition to inhomogeneity? Or some other unobvious conservation laws emerge? Even if this were not the case, ergodicity is still not a guaranteed property, so the connection between conservation laws and stability remains true only in the more restricted sense we emphasize here.

While Morinaga's sufficiency proof \cite{Morinaga:2021vmc} is quite involved, its thrust suggests a significant simplification in that the guaranteed unstable modes have a $\bk$ that points in a direction $\bv_c$ towards a crossing line of $G_\bv$, i.e., $G_{\hat\bk}\simeq0$, are nearly luminal, $\omega\simeq|\bk|$, and have a weak growth rate, being at the edge of an unstable region.

So the second thrust of our investigation is connected with a resonant exchange of energy between flavor waves and individual neutrinos, a mechanism particularly relevant for weak instabilities. To show this resonance effect, we start from a stable configuration and construct the theory of fast collective oscillations in a linear-response framework, where they arise as the response of the neutrino gas to an external flavor field. The usual approach of looking for eigenmodes oscillating in time, analogous to the first attempt by Vlasov to solve his own equation for collisionless plasmas \cite{Vlasov:1945}, is in principle correct, but only if all eigenmodes are kept, including the continuum of the often-forgotten Case-van Kampen modes \cite{VanKampen:1955,Case:1959}. This makes for a cumbersome approach of solving an actual initial-value problem. In the linear-response framework, following Landau \cite{Landau:1946jc}, one can show much more easily that fast collective oscillations are Landau damped. The origin of Landau damping is related to the streaming nature of the system; for a field depending on time and space as $\psi_\bv e^{-i(\omega t-\bk\cdot\br)}$, the Vlasov operator on the lhs of the QKE vanishes when $\omega-\bv\cdot\bk=0$. Thus, there are particles that are resonant with the collective wave if it is subluminal (space like), meaning $\omega^2<\bk^2$. In the well-studied plasma case, this situation means that energy can be transferred from the collective mode to individual electrons, Landau-damping the wave. This effect can also be understood as Cherenkov absorption of plasma waves by the electrons. This analogy reveals that the same happens in the flavor case, where neutrinos resonating with a flavor wave can absorb its energy and damp it. It also holds the key for an intuitive understanding of how instabilities arise; their appearance means that the process gets reversed, with neutrinos compulsively emitting (rather than absorbing) Cherenkov flavor waves.

We will use this basic intuition to prove that angular crossings {\em necessarily} lead to instabilities, simplifying  Morinaga's proof of sufficiency~\cite{Morinaga:2021vmc} and, more importantly, provide an intuitive physical interpretation. We will show that this result follows at once from the resonant picture. The sign of the energy exchange due to Cherenkov emission depends on the sign of the distribution $G_\bv$. If a non-crossed distribution is slightly distorted so as to have a crossing line, marking the boundary of a region where the neutrino distribution is flipped in sign, the flavor waves that are primarily resonant with the neutrinos in the flipped region will gain energy rather than lose it from the Cherenkov process. Those waves correspond to unstable modes. As a benefit, we immediately learn that neutrinos in the flipped region are the ones that feel the effect of the instability the strongest -- since they are the ones that caused it in the first place -- providing an intuitive explanation for the ``removal of the angular crossing'' that was found numerically~\cite{Nagakura:2022kic,Zaizen:2022cik} and that we recently proved theoretically based on quasi-linear theory~\cite{Fiorillo:2024qbl}. On the basis of our simple physical ideas, we construct a full proof of Morinaga's result, although admittedly, the rigorous proof also involves some technicalities.

We structure our work by first reconsidering in Section~\ref{sec:conservation_laws} the conventional viewpoint on collective oscillations in light of conservation laws, showing that the necessity of an angular crossing from instability is directly implied. In Section~\ref{sec:linear_response}, we develop the fast flavor evolution as the response to an external field, introducing the notion of flavor susceptibility. In Section~\ref{sec:initial_condition}, we show how the flavor susceptibility is recovered also in terms of an initial-value formulation of the problem. We also introduce quantitatively the notion of resonant exchange of energy which lies behind the instability. In Section~\ref{sec:angular_crossing}, we apply these ideas to formulate a physical argument and a mathematical proof for the connection between angular crossings and instabilities. Finally, in Section~\ref{sec:discussion}, we summarize our results. In Appendix~\ref{sec:Landau}, as a a pedagogical exercise and historical excursion, we recap the corresponding development in the context of plasma waves. In Appendix~\ref{app:lepton_number_conservation}, we provide technical details about the symmetry argument leading to the necessity of crossings for the instability of fast, slow, and collisional evolution.

\section{Collective flavor oscillations and conservation laws}
\label{sec:conservation_laws}

\subsection{Equations of motion}

In this section, we review the conventional QKE for collective flavor oscillations, and discuss the concept of stability from the point of view of the conservation laws of the system. As we have recently shown~\cite{Fiorillo:2024fnl}, the QKEs follow immediately if we consider the renormalized energy of massless neutrinos with momentum $\bp$ as
\begin{equation}
    {\sOm_\bp}(x)=|\bp|+\sqrt{2}\GF \int d\Pi'\,v\cdot v'\,\varrho_{\bp'}(x),
\end{equation}
where $\GF$ is Fermi's constant, $d\Pi=d^3\bp/(2\pi)^3$ the phase-space element, $v^\mu=p^\mu/|\bp|$ the velocity four-vector, and $p^\mu=(|\bp|,\bp)$ the four-momentum; we denote by a dot the covariant four-product. Under the integral, we have left out a term proportional to $\mathrm{Tr}(\rho_{\bp'})$ that appears starting from first principles \cite{Fiorillo:2024fnl}. However, provided that the system is not strongly inhomogeneous, this term has negligible impact and is not needed for the consistency of the equations of motion (EOMs), although energy conservation is no longer fulfilled. Finally, with the space-time coordinates $x=(t,\br)$, the density matrix of neutrinos with momentum $\bp$ is
parameterized as
\begin{equation}
    \varrho_\bp
    =\frac{1}{2}\bigl({\rm Tr}\varrho_\bp+\vec{P}_{\bp}\cdot\vec{\sigma}\bigr)
    =\frac{1}{2}\begin{pmatrix}
        n_\bp+P^z_\bp & P^x_\bp-iP^y_\bp \\
        P^x_\bp+iP^y_\bp & n_\bp-P^z_\bp
    \end{pmatrix}
\end{equation}
with the neutrino phase-space density $n_\bp$ and polarization vector $\vec{P}_\bp$. For the off-diagonal term that represents the field of flavor coherence, we use $\psi_\bp=P^x_\bp+iP^y_\bp$. In the previous literature, often the upper-right element $P^x_\bp-iP^y_\bp$ was used instead, explaining certain sign differences in our equations. The directions $\{x,y,z\}$ in flavor space should not be confused with the same-named directions in coordinate space. 

Neutrinos with this renormalized dispersion relation bear some conceptual resemblance to the Fermi-liquid picture of a system of interacting fermions, where one also regards the elementary excitations as individual quasi-particles with a renormalized dispersion relation. Indeed our collective oscillations of flavor share a similar origin with collective oscillations -- of density, spin, or any other quantity -- in Fermi liquids. It is important to stress, however, the different reason: in Fermi liquids, the quasi-particle picture applies because degeneracy ensures that the scattering rate is always much smaller than the quasi-particle energy near the Fermi surface. In the flavor case, the scattering rate of individual neutrinos is intrinsically small relative to flavor conversion scales.

The density matrix evolves according to the often-derived standard quantum kinetic equations \cite{Dolgov:1980cq, Rudsky, Sigl:1993ctk, Sirera:1998ia, Yamada:2000za, Vlasenko:2013fja, Volpe:2013uxl, Serreau:2014cfa, Kartavtsev:2015eva, Fiorillo:2024fnl, Fiorillo:2024wej}
\begin{equation}
    \partial_t \varrho_\bp + \frac{1}{2}\left\{\bpartial_\bp \sOm_\bp,\bpartial_\br \varrho_\bp\right\}- \frac{1}{2}\left\{\bpartial_\br \sOm_\bp,\bpartial_\bp \varrho_\bp\right\}=i[\varrho_\bp,\sOm_\bp].
\end{equation}
The commutator is the quantum generalization of the usual Liouville equation, while the anticommutator is the usual quantum evolution driven by the one-particle Hamiltonian~$\sOm_\bp$. We have recently shown that this conventional kinetic equation not only reproduces the traditional EOM, but also conserves energy to all orders \cite{Fiorillo:2024fnl}. To recover the standard form, we assume that there is no net transport of neutrinos, allowing us to concentrate on the trace-free part. In the massless limit, any dependence on energy $E$ drops out, so we may assume appropriately phase-space integrated quantities with all dimensionful factors absorbed in the units of space and time, immediately leading to
\begin{equation}
   (\partial_t+\bv\cdot\bpartial_\br)\vec{P}_\bv=
    v\cdot\partial \vec{P}_\bv=\int d^2\bv'\, v\cdot v'\,
    \vec{P}_{\bv'}\times \vec{P}_\bv,
\end{equation}
where only the direction of motion $\bv$ remains as a variable to denote different modes and the remaining phase-space integral $\int d^2\bv$ is over the unit sphere of directions. Notice that the length of the polarization vectors is automatically conserved along the flow due to the precession nature of the rhs. 

Usually we consider a system for which the density matrix is nearly diagonal, with $P^z_\bv\big|_{t=0}= G_\bv$ as initial value and $|\psi_\bv|\ll |G_\bv|$ initially. In the off-diagonal terms of this EOM, we can set $P^z_\bv\simeq G_\bv$ in the linear regime because $P^z_\bv$ changes quadratically in the perturbation $\psi_\bv$. We thus recover
\begin{equation}\label{eq:eom_transverse}
    v\cdot\partial\psi_\bv=-i\int d^2\,\bv'v\cdot v'\,
    \bigl(G_\bv \psi_{\bv'}-G_{\bv'}\psi_\bv\bigr), 
\end{equation}
which indeed reproduces Eq.~\eqref{eq:neutrino_eom}. It is convenient to introduce a compact notation
\begin{equation}
    G^\mu=\int d^2\bv\,G_\bv v^\mu,
\end{equation}
and similarly for $\vec{P}^\mu$ and $\psi^\mu$, leading to the linearized EOMs
\begin{equation}\label{eq:EOM-linear}
    v\cdot\partial\psi_\bv=i(G\cdot v\, \psi_\bv-\psi\cdot v\, G_\bv).
\end{equation}
In this form, the EOMs are more compact than in the previous literature, where the field of flavor coherence was usually defined in the normalized form $S_\bv=\psi^*_\bv/G_\bv$. In terms of $S_\bv$, the EOM involves an integral on the rhs and the opposite sign, although final results are, of course, the same.

\subsection{Normal modes}

What to make of the collisionless linearized Vlasov-type EOM of Eq.~\eqref{eq:EOM-linear} depends on which problem one intends to solve. For any strong flavor conversion effects to occur, the appearance of exponentially growing modes of the field of flavor coherence is central. To find these, one looks for solutions $\psi_\bv(t,\br)$ in the form of normal modes $\psi_\bv(\Omega,\bK)\,e^{-i\Omega t+i \bK\cdot \br}=\psi_\bv(K)\,e^{-iK \cdot x}$, where for notational simplicity we denote the normal-mode eigenfunction with the same letter $\psi_\bv$. Fast flavor effects are expected to occur on small scales, so one imagines an essentially homogeneous ensemble evolving in time and so one studies spatial Fourier modes (real $\bK$) and asks if their time evolution is harmonic (real $\Omega$) or growing/damped with a complex $\Omega$, depending on $G_\bv$. Mathematically, in analogy to plasma physics, we assume the system to be spatially infinite. In this idealization, the problem of boundary conditions does not even need specification, and the only assumption is that the unperturbed neutrino properties change over scales much larger than the ones of fast conversions. Searching for normal modes is a different exercise from solving the linear EOM in Fourier space because a mode with complex $\Omega$ is not a Fourier mode, but rather appears in the Laplace transform as we will see in formal detail later.

Substituting the normal-mode ansatz in the EOM immediately reveals that they must satisfy
\begin{equation}\label{eq:normal_mode_condition}
    k\cdot v\, \psi_\bv=\psi\cdot v\, G_\bv,
\end{equation}
where we have introduced $k^\mu=K^\mu+G^\mu$ with components $k=(\omega,\bk)$, absorbing the neutrino matter effect on each other in the usual way in the definition of frequency and momentum. The usual next step is to divide by $k\cdot v$, multiply by $v^\mu$, and integrate over $d^2\bv$ to obtain the consistency equation
\begin{equation}\label{eq:wrong_chi}
    \psi^\mu={\chi}^\mu_\nu \psi^\nu
    \quad\hbox{with}\quad
    {\chi}^\mu_\nu=\int d^2\bv \frac{G_\bv v^\mu v_\nu }{k\cdot v}.
\end{equation}
Nontrivial solutions require ${\rm det}(g^{\mu\nu}-\chi^{\mu\nu})=0$ and this would be the condition that determines the normal modes \cite{Izaguirre:2016gsx, Capozzi:2017gqd}. However, in reaching this condition there is a critical passage that is generally incorrect, namely the division by $k\cdot v$. If $k$ is real and space like, such that $\omega^2-\bk^2<0$, there will be certain values of $\bv$ for which $k\cdot v=0$, making the integral ill-defined. The only reason one does not usually worry about this problem is that the unstable normal modes, with $\mathrm{Im}(\Omega)=\mathrm{Im}(\omega)>0$, do not suffer from this problem. So if the only purpose of this strategy is to diagnose the existence of such modes, this procedure leads to the correct answer. However, solving a concrete initial-value problem by such means is generally impossible.

One can still use the normal modes identified by Eq.~\eqref{eq:normal_mode_condition} to determine the evolution of a given initial condition, but only by understanding them more carefully. Any initial condition can be decomposed as a sum of eigenmodes $\psi^{(i)}_\bv$ and then carried forward in time as $e^{-i k^{(i)}\cdot x}$, leading in principle to a solution of the initial-value problem, assuming one has a complete set of linearly independent eigenmodes. However, Eq.~\eqref{eq:wrong_chi} only allows one to find those eigenmodes with $\mathrm{Im}(\Omega)\neq 0$, and the superluminal eigenmodes with $\mathrm{Im}(\Omega)=0$ and $\omega^2>\bk^2$, for which $k\cdot v$ never vanishes. 

What about the subluminal eigenmodes with $\mathrm{Im}(\Omega)=0$ and $\omega^2\leq \bk^2$?
For them, we return to the condition Eq.~\eqref{eq:normal_mode_condition} and seek the solution in the form
 \begin{equation}
     \psi_\bv=\delta(k\cdot v)+f^k_\bv,
 \end{equation}
where the function $f^k_\bv$ is chosen to vanish when $k\cdot v=0$. Substitution in Eq.~\eqref{eq:normal_mode_condition} reveals
 \begin{equation}
     f^k_\bv=\frac{\psi\cdot v\,G_\bv}{k\cdot v}.
 \end{equation}
We now have a definite prescription on how to interpret the $k\cdot v$ division; since $f^k_\bv$ by assumption vanishes when $k\cdot v=0$, we should interpret it in the principal-value sense. Multiplying by $v^\mu$ and integrating, we now find that for these eigenmodes the self-consistency condition reads
 \begin{equation}\label{eq:new-selfconsistency}
     \psi^\mu={\chi}^\mu_\nu \psi^\nu+\int d^2\bv\,\delta(k\cdot v)\,v^\mu.
 \end{equation}
Crucially, Eq.~\eqref{eq:new-selfconsistency} is \textit{not} a homogeneous equation for $\psi^\mu$ and therefore does not provide a dispersion law. There are solutions for any value of $k$ -- provided that $\mathrm{Im}(\Omega)=0$ and $\omega^2<\bk^2$. Thus, there is a continuous set of eigenmodes that were missing from the standard approach. These eigenmodes, which in the fast flavor context were identified in Ref.~\cite{Capozzi:2019lso}, were much earlier discovered in the plasma-physics context  (see Appendix~\ref{sec:Landau}) and are known as Case-van Kampen modes~\cite{VanKampen:1955wh, case1959plasma, Sagan:1993es}. The connection between the ``non-collective fast flavor modes'' of Ref.~\cite{Capozzi:2019lso} and the Case-van Kampen modes of plasma physics was first made in Ref.~\cite{Fiorillo:2023mze} in the context of homogeneous fast flavor conversions.

\subsection{Stability from conserved quantities}

A different perspective on possible strong flavor evolution derives from the global properties of the system. In most cases, the appearance of unstable collective oscillations around a state means that it is not at a minimum of its energy. This is the case, for example, for the close analog system of a Fermi liquid, where unstable collective oscillations appear as soon as a ground state with lesser energy than the standard Fermi sphere appears~\cite{pomeranchuk1959stability}. The reason for this minimum-energy principle is simple: if the system is at a minimum of energy, there is no place to go without violating energy conservation. The same could be said about a maximum energy configuration, although such configurations are usually of lesser interest in many-body physics, where the system is typically considered to be coupled to a thermal bath with zero temperature. In any case, it is clear that an isolated system in a true extremum point of its energy functional is stable because there are no other configurations it can go to. If the system is not at such an extremum, it has freedom to move and usually will be unstable unless it is protected by other symmetries.

For our flavor system, there are actually two fundamental conservation laws, one being its total energy
\begin{equation}
    \mathcal{U}=\int d^3\br\left(\frac{\vec{P}_\mu\cdot\vec{P}^\mu}{4}
    +\int d\Pi\, |\bp|\, n_\bp \right).
\end{equation}
The other is the conservation of lepton number $\int d\Pi\, d^3\br\,\varrho_\bp(\br)$, which separately implies the conservation of $\int d^3\br\, n^0$ (the total number of neutrinos) and $\int d^3\br\, \vec{P}^0$ (the $z$-component of this quantity being the total lepton number of neutrinos). In fact, the conservation law for the $z$-component $\int d^3\br P^0_z$ is quite general, and it holds for the fast, slow, and collisional flavor instability; we discuss this explicitly in Appendix~\ref{app:lepton_number_conservation}. If we interpret the polarization vectors $\vec{P}_\bv$ as interacting spins, the conservation of $\int d^3\br\,\vec{P}^0$ is analogous to the conservation of total angular momentum. The stability argument must therefore be enforced as follows: if the system is in an extremum of energy or angular momentum, it must be stable. Initialized with an arbitrary distribution $G_\bp$, the system certainly is not in an extremum of the energy because one can always lower or increase the neutrino kinetic energies without violating Pauli's principle, since neutrinos are not degenerate. One could think that this kinetic energy would not have anything to do with the spin dynamics, since it does not appear in the EOM. However, as we have recently shown~\cite{Fiorillo:2024fnl}, this is not true for inhomogeneous systems. Even a weakly inhomogeneous flavor structure produces a weak potential gradient that exerts a force on the neutrinos. Thus, the fact that the kinetic energy is neither in a minimum nor a maximum is enough to understand that energy conservation cannot enforce stability.

Another obvious possibility is the conservation of angular momentum $\int d^3\br\,\vec{P}^0$. If $G_\bv$ has the same sign everywhere (no crossing), all spins are initially aligned and thus obviously in a maximum of total angular momentum. There is no place to go without violating angular momentum conservation. Hence, we immediately see that an angular crossing is necessary for an instability, a conclusion otherwise reached with much more effort~\cite{Morinaga:2021vmc,Dasgupta:2021gfs}. As mentioned earlier, this symmetry argument was very recently presented in similar form in Ref.~\cite{Johns:2024bob}. This conclusion is even more general and applies to any system for which the $z$-component of the total angular momentum is conserved. In particular, this includes the case of slow flavor conversions, in which the mass axis is usually taken to coincide with the flavor axis due to matter effects and hence oscillations correspond to precession around the $z$-axis. And it also applies to the collisional instability if only the components transverse to the flavor direction are damped and therefore the $z$-component is conserved. An explicit discussion of these cases is deferred to Appendix~\ref{app:lepton_number_conservation}. 

If an angular crossing is present, one would expect the system to be unstable because it can reach other configurations. However, such a conclusion is premature without proving first the absence of other conserved quantities that may be less obvious than angular momentum. Thus far, Morinaga's proof of sufficiency of a crossing \cite{Morinaga:2021vmc}, technically only applicable to the fast flavor case, is the only available indication that this is not the case and the system is not protected by some other invisible symmetry. Without this (or our later) proof of sufficiency, the absence of angular-momentum protection does not guarantee instability. \textit{After} that proof has been executed, one can conclude that the system will certainly visit other phase-space configurations, although it remains speculative whether it will visit \textit{any} allowed configuration or if it will do so with equal probability, as implied by the ergodicity assumption of Ref.~\cite{Johns:2024bob}.

\section{Linear-response theory of flavor waves}\label{sec:linear_response}

\subsection{General idea}

The conventional view on flavor instabilities is that certain initial conditions for the neutrino flavor field, when seeded with a small perturbation, will change dramatically. We call this a picture with perturbed initial conditions. The rationale is that, even though one initializes the system with neutrinos in their unperturbed flavor state, the mass term will provide some intrinsic deviations from the flavor basis, suppressed both by the small mass and by the small effective mixing angle in matter. Due to the exponential growth of the instability, one expects the evolution to be mostly insensitive to the exact nature of the initial seeding. In this sense, one is not solving a specific initial-value problem and not providing specific perturbed initial conditions.

Therefore, we may entertain an alternative viewpoint, which gives new insights on the meaning of instability. Instead of perturbed initial conditions as the trigger for the instability, we consider unperturbed initial conditions, but that the neutrino gas is subject to a small external flavor field. Hence, we imagine that the renormalized neutrino energy can be written as ${\sf\Omega}_\bp=|\bp|+\varrho\cdot v+\varrho_\mathrm{ext}\cdot v$, where $\varrho^\mu=\int d\Pi\,\varrho_\bp v^\mu$, and we have introduced an external density matrix $\varrho_\mathrm{ext}^\mu$, which we can write as
\begin{equation}
    \varrho_{\mathrm{ext},\mu}=\frac{1}{2}\begin{pmatrix}
        0 & \psi^*_{\mathrm{ext},\mu} \\
        \psi_{\mathrm{ext},\mu} & 0
    \end{pmatrix}.
\end{equation}
We can call the total off-diagonal field felt by each neutrino $\Psi^\mu=\psi^\mu+\psi_\mathrm{ext}^\mu$, where we recall that $\psi^\mu=\int d^2\bv\,v^\mu \psi_\bv$ and $G^\mu=\int d^2\bv\,v^\mu G_\bv$.

This apparently artificial way of triggering the fast flavor instability is actually the only physical one. If all neutrinos are produced in flavor eigenstates, deviations from the flavor axis do not initially exist, but rather come from the mass term that is neglected in the fast flavor context. It would appear in the effective neutrino energy as
\begin{equation}
    \delta{\sf\Omega}_\bp={\sf\Omega}^\mathrm{vac}_\bp+{\sf\Omega}^\mathrm{mat},
\end{equation}
with
\begin{equation}
   {\sf \Omega}^\mathrm{vac}_\bp=\frac{\delta m^2}{2|\bp|}{\sf B}
   \quad\hbox{and}\quad
   {\sf\Omega}^\mathrm{mat}=\sqrt{2}\GF\,\frac{n_e}{2} \sigma_z,
\end{equation}
where
\begin{equation}
    {\sf B}=\frac{1}{2}\begin{pmatrix}
        \cos(2\theta) & \sin(2\theta)\\
        \sin(2\theta) & -\cos(2\theta)
    \end{pmatrix}
\end{equation}
and $\theta$ is the vacuum mixing angle, $\delta m^2$ the squared mass splitting, and we include the matter refraction term, where $\sigma_z$ is the third Pauli matrix and $n_e$ is the electron number density; we assume it to be homogeneous and constant. Inhomogeneities in the matter density profile might directly couple to inhomogeneous perturbations in the flavor evolution~\cite{Airen:2018nvp}; we do not explore this topic further here. Since matter refraction usually exceeds all other terms, we can remove it as usual by going to a frame co-rotating around the $z$-direction (the flavor direction), thus redefining the density matrix $\varrho_\bp \to e^{-i\int {\sf\Omega}^\mathrm{mat}_\bp dt}\varrho_\bp e^{i\int {\sf\Omega}^\mathrm{mat}_\bp dt}$, after which the mass Hamiltonian is changed to
\begin{equation}
    {\sf \Omega}_\bp^\mathrm{vac}\to e^{i\int {\sf\Omega}^\mathrm{mat}_\bp dt} {\sf \Omega}_\bp^\mathrm{vac} e^{-i\int {\sf\Omega}^\mathrm{mat}_\bp dt}=\frac{\delta m^2}{4|\bp|}\begin{pmatrix}
        \cos(2\theta) & \sin(2\theta) e^{i\sqrt{2}\GF\int n_e dt}\\
        \sin(2\theta) e^{-i\sqrt{2}\GF\int n_e dt} & -\cos(2\theta)
    \end{pmatrix}.
\end{equation}
We can now see that the off-diagonal part of this matrix acts exactly as the external field we introduced as a trigger for the instability, with a time dependence driven by the external matter profile. 

In this linear-response approach, the external field is truly a probe that disturbs the system to examine its response. Collective oscillations appear as resonances, in the same way as if we tickle a harmonic oscillator close to its natural frequency. We mostly discuss stable systems and only later generalize to unstable ones.  

Including the external field, the EOM is identical to the conventional form of Eq.~\eqref{eq:EOM-linear}, except that the self-consistent field $\psi^\mu$ is augmented on the rhs by the external field $\psi_\mathrm{ext}^\alpha$ so that
\begin{equation}
    v_\alpha (i\partial^\alpha+G^\alpha)\psi_\bv=v_\alpha \Psi^\alpha G_\bv.
\end{equation}
In principle, the external field could be applied as an impulse at some instant. However, a much more convenient approach is to imagine that a harmonic form is inserted infinitely slowly so that $\psi_\mathrm{ext}^\alpha \propto e^{-i\Omega t+i\bK\cdot \bx+\epsilon t}$, where $\epsilon\to 0$ is positive. Since the equations are linear, it is clear that the neutrinos respond with a field $\psi_\bv\propto e^{-i\Omega t+i\bK\cdot\bx+\epsilon t}$. Inserting this solution in the EOM reveals 
\begin{equation}\label{eq:response_field}
    \psi_\bv=\frac{G_\bv v_\alpha \Psi^\alpha}{k^\mu v_\mu+i\epsilon}.
\end{equation}
Multiplying with $v^\beta$ and integrating over velocities, we find the response field as
\begin{equation}
    \psi^\alpha=\int d^2\bv\, \psi_\bv v^\alpha= \chi^{\alpha\beta}\Psi_\beta,
\end{equation}
where
\begin{equation}\label{eq:susceptibility}
    \chi^{\alpha\beta}=\int d^2\bv\, \frac{G_\bv v^\alpha v^\beta}{k^\mu v_\mu+i\epsilon}
\end{equation}
is the flavor response function, which describes the response of the system to the \textit{total} field $\Psi$. We will later introduce the susceptibility as the response to the external field $\psi_\mathrm{ext}^\mu$ alone. The flavor response function of Eq.~\eqref{eq:susceptibility} is analogous to Eq.~\eqref{eq:wrong_chi}, except for the explicit prescription of how to integrate around the pole, which here arises from introducing the perturbation slowly in time.

\subsection{Flavor response function}

Before turning to self-consistent solutions, it is instructive to understand the meaning of the flavor response function. It describes the response of the system to the total field. If the forcing field is much larger than the response $\psi_\mathrm{ext}^\mu\gg \psi^\mu$, then we may approximate $\psi^\mu=\chi^{\mu}_{\nu} \Psi^\nu\simeq \chi^\mu_\nu \psi_\mathrm{ext}^\nu$. Such a condition will naturally happen if the external field is not resonant with a collective mode. A key insight from Eq.~\eqref{eq:susceptibility} is that the flavor response function is complex; in the limit $\epsilon\to 0$
\begin{equation}
    \chi^{\alpha\beta}=\fint d^2\bv\,\frac{G_\bv v^\alpha v^\beta}{k^\mu v_\mu}-i\pi \int d^2\bv\, G_\bv v^\alpha v^\beta \delta(k^\mu v_\mu).
\end{equation}
It has acquired an imaginary part, which receives contributions only from neutrinos moving \textit{in phase} with the wave; we will dub these \textit{resonant neutrinos}. What does a complex response function mean? At first sight, it only means that the response does not move in phase with the external field. However, from the general theory of linear response, we know that there is a deeper meaning; the imaginary part of the response function implies that there is irreversibility and ultimately dissipation of energy.

In the present case, energy is indeed not conserved because the external field varies in time. The energy variation comes entirely from the time derivative of the external field
\begin{equation}
    \frac{dE}{dt}=\int d\Pi\,\mathrm{Tr}\left(\frac{d\varrho_\mathrm{ext}^\mu}{dt}v_\mu \varrho_\bp\right).
\end{equation}
Inserting the explicit form of $\rho_\mathrm{ext}^\mu$ yields
\begin{equation}
    \frac{dE}{dt}=\int d^2\bv\,\frac{i\Omega}{4}v^\mu \left(\psi_\bv \psi^*_{\mathrm{ext},\mu}-\psi_\bv^*\psi_\mathrm{ext}^\mu\right)
    =\frac{i\Omega}{4}\left(\psi^\mu \psi^*_{\mathrm{ext},\mu}-\psi^*_\mu \psi^{\mathrm{ext},\mu}\right).
\end{equation}
We now express $\psi^\mu$ in terms of $\Psi^\mu$ using the flavor response function and finally obtain
\begin{equation}
    \frac{dE}{dt}=-\frac{\Omega}{2} \psi^*_{\mathrm{ext},\mu} \psi_{\mathrm{ext},\nu} \mathrm{Im}(\chi^{\mu\nu}).
\end{equation}
This is the familiar result that the imaginary part of the susceptibility describes energy dissipation; specifically, if this expression is positive, the system absorbs energy from the field, and vice versa. (We recall that presently we work in the limit of small response where our flavor response function coincides with the susceptibility to the external field alone.) If the system absorbs energy in this way, the field must be damped; this effect is generally known as Landau damping.

As we have seen, the imaginary part of the susceptibility is entirely determined by the resonant neutrinos moving in phase with the wave. This simple physical conclusion deserves an equally simple explanation. Let us assume that the external field varies with the four vector $K^\mu$. A neutrino with velocity $v^\alpha$ moves along the trajectory $\br=\bv t$; along this trajectory, we can express the evolution as the total derivative $d/dt=v_\alpha \partial^\alpha=\partial_t+\bv\cdot\partial_\br$. Hence, the off-diagonal component of the density matrix for this particular neutrino follows a Schr\"odinger equation
\begin{equation}
    i\frac{d\psi_\bv}{dt}+v\cdot G \psi_\bv=G_\bv\, v\cdot \Psi\, e^{-i(\Omega-\bK\cdot\bv+i\epsilon)\, t}.
\end{equation}
If we introduce the phase factor
\begin{equation}
    \psi_\bv=\overline{\psi}_\bv e^{iv\cdot G\, t}
\end{equation}
we find immediately an equation for $\overline{\psi}_\bv$
\begin{equation}\label{eq:evo_line_sight}
    i\frac{d\overline{\psi}_\bv}{dt}=G_\bv\,v\cdot \Psi\,e^{-i(k\cdot v+i\epsilon) t},
\end{equation}
where again $k^\mu=K^\mu+G^\mu$. Usually the rhs is a rapidly oscillating function and hence the neutrino will sometimes gain and sometimes lose energy from the external field. On the other hand, resonant neutrinos for which $k\cdot v=0$ always see a constant forcing term, and therefore constantly exchange energy with the external field. The external field energy is thus dissipated into the flavor precession of individual resonant neutrinos. 

We notice here that neutrinos moving in resonance with the wave feel a ``force'' term on the rhs of Eq.~\eqref{eq:evo_line_sight} proportional to $G_\bv$. Hence, neutrinos with opposite sign of $G_\bv$ feel an effective forcing term with opposite sign. In turn, the opposite sign implies opposite rates of change of the energy; according to the sign of $G_\bv$, neutrinos can draw or give energy to the mode. As we will see, this has a fundamental connection with the role of an angular crossing in the instability formation. The geometrical reason is simple; if the polarization vector of the neutrino points downward, its precession around the nearly vertical collective field will go in the opposite sense than if it were pointing upward. 

\subsection{Flavor susceptibility}

We will now discuss the response of the system to the external field alone $\psi^\mu=\alpha^{\mu\nu} \psi^\mathrm{ext}_\nu$, which we call flavor susceptibility\footnote{Regarding terminology, in electrodynamics it is common to call \textit{susceptibility} the response of the system to the total electric field, similar to what we have called the flavor response function. The reason is that in electrodynamics one usually does not distinguish between the response to an external or internal electric field, while conventionally one distinguishes external charges and currents from internal ones. In systems in which external fields are usually adopted, e.g.\ Fermi liquid theory~\cite{lifshitz2013statistical}, the susceptibility is more properly defined as the response to the external field alone.}, such that
\begin{equation}
    \alpha^\mu_\nu=(g^{\mu\alpha}-\chi^{\mu\alpha})^{-1}\chi^\nu_\alpha,
\end{equation}
where by $A^{-1}$ we mean the inverse of the matrix $A$. The existence of self-consistent collective oscillations is signaled by a diverging flavor susceptibility, which means that there is a resonance. Such a divergence appears when the matrix in parenthesis cannot be inverted, leading to the dispersion relation
\begin{equation}\label{eq:dispersion_causal}
    \Phi_{\omega,\bk}=\mathrm{det}[g^{\mu\nu}-\chi^{\mu\nu}]=0.
\end{equation}
At first sight, this looks identical to the dispersion relation for the normal modes, yet it differs in a fundamental aspect, namely that the integral must be performed with the $i\epsilon$ prescription. We will usually refer to this as the Landau prescription, as it was originally introduced in the context of the Vlasov equation for plasma waves by Landau in the seminal Ref.~\cite{Landau:1946jc} -- see Appendix~\ref{sec:Landau} for a brief recap.

We can now distinguish two different cases, beginning with \textbf{superluminal modes}, defined by $\omega^2>\bk^2$ with real $\omega$. The tensor $\chi^{\mu\nu}$ remains real because $\omega-\bk\cdot\bv$ never vanishes. Thus, one can expect that there are collective undamped modes; in this case, the flavor susceptibility diverges close to these frequencies. This is simply the usual behavior of a oscillator, in this case with many degrees of freedom, responding to a force resonant with its natural frequency. 

The second case refers to \textbf{subluminal modes}, defined by $\omega^2>\bk^2$ for real $\omega$. The tensor $\chi^{\mu\nu}$ develops an imaginary part, and thus one does not generally expect solutions of the dispersion relation with real $\omega$. For a stable system, the only possibility for collective modes is that the dispersion relation Eq.~\eqref{eq:dispersion_causal} has solutions with a negative imaginary part; one of them shall be called $\overline\omega=\overline\Omega+G^0$ for some unspecified wavevector and we write explicitly $\overline{\Omega}=\overline{\Omega}_{\rm R}+i\overline{\Omega}_{\rm I}=\overline{\Omega}_{\rm R}-i|\overline{\Omega}_{\rm I}|$ in terms of the real and imaginary parts. If the system is probed \textit{exactly} by a monochromatic field $\psi^\mu_\mathrm{ext}=\tilde{\psi}^\mu_{\mathrm{ext},\Omega} e^{-i\Omega t}$ with a real $\Omega$ close to $\overline{\Omega}_{\rm R}$, and assuming $|\Omega_{\rm I}|\ll |\Omega_{\rm R}|$, then the flavor susceptibility close to the pole has a characteristic behavior $\alpha^{\mu\nu}=\alpha^{\mu\nu}_0/(\Omega-\overline{\Omega}_{\rm R}+i|\overline{\Omega}_{\rm I}|)$. Again, this is completely analogous to a simple damped oscillator; if the latter is excited by a monochromatic force, it performs undamped oscillations, with the external field providing the energy to overcome damping. If the external field is not monochromatic, but has a frequency distribution $\psi^{\mu}_\mathrm{ext}=\int d\Omega\,\tilde{\psi}^\mu_{\mathrm{ext},\Omega} e^{-i\Omega t}$, the response will be
\begin{equation}\label{eq:late_time_integral}
        \psi^\mu=\int d\Omega\, \alpha^\mu_\nu \tilde{\psi}^\nu_{\mathrm{ext},\Omega} e^{-i\Omega t}\simeq \int d\Omega\,\alpha^\mu_{0,\nu}\tilde{\psi}^\nu_{\mathrm{ext},\Omega}\,\frac{e^{-i\Omega t}}{\Omega-\overline{\Omega}_{\rm R}+i|\overline{\Omega}_{\rm I}|},
\end{equation}
where we approximate the response close to the pole. This integral at late times behaves as $e^{-i\Omega_{\rm R} t} e^{-|\Omega_{\rm I}| t}$, and thus is exponentially damped (the non-pole contributions lead to subdominant contributions at late times, unless they include other kinds of singularities as we discuss in Sec.~\ref{sec:analytic_properties}). The damping is generally known as Landau damping; even though the system does not have any scattering mechanism, and its EOM are invariant under time reversal, the damping arises from the continuous degrees of freedom. Information and energy is lost to the individual velocity modes, which rapidly decohere from one another, damping the collective field.

Notice that the dispersion relation for normal modes, conventionally adopted in the fast flavor literature, does not predict Landau damping. This behavior does not represent a normal mode, but rather an asymptotic behavior at late times coming from a superposition of the continuous normal modes which are individually not damped, but rapidly decohere relative to each other. We also notice that for this stable configuration the poles of the flavor susceptibility are all in the lower half-plane of the complex frequency $\Omega$, which is the condition that guarantees causality and in turn the well-known Kramers-Kronig relations for the refractive index in the context of photon propagation \cite{kramers1928diffusion,deL.Kronig:26}. 

Let us finally turn to the more nuanced case of unstable systems. The initial procedure of applying an external field infinitely slowly becomes now unrealistic. If the dispersion relation has a pole at $\overline{\Omega}=\overline{\Omega}_{\rm R}+i\overline{\Omega}_{\rm I}$ (we assume $\overline{\Omega}_{\rm I}>0$ is the maximum growth rate of the system), then any infinitesimal perturbation in the application of the field would grow much faster than the field itself. Hence, to apply consistently an external field approach to an unstable system, we must insert it as $e^{-i\Omega t}=e^{-i\Omega_{\rm R} t+(\overline{\Omega}_{\rm I}+\epsilon)t}$, so that its insertion is faster than the maximum growth rate. In this case, when we apply a non-monochromatic field, the integral in Eq.~\eqref{eq:late_time_integral} takes a formally similar expression
\begin{equation}\label{eq:late_time_integral-2}
        \psi^\mu=\int d\Omega\,\alpha^\mu_\nu \tilde{\psi}^\nu_{\mathrm{ext},\Omega} e^{-i\Omega t}\simeq \int d\Omega\,\alpha^\mu_{0,\nu}\tilde{\psi}^\nu_{\mathrm{ext},\Omega} \frac{e^{-i\Omega t}}{\Omega-\overline{\Omega}_{\rm R}-i\overline{\Omega}_{\rm I}},
    \end{equation}
except that the integral over $\Omega$ is now done for $\Omega=\Omega_{\rm R}+i(\overline{\Omega}_{\rm I}+\epsilon)$. The late-time behavior of this integral is now of course proportional to $e^{-i\overline{\Omega}_{\rm R} t +\overline{\Omega}_{\rm I} t}$, exponentially growing in time as expected. The necessity of this procedure is already evident from the fact that otherwise the flavor susceptibility in the upper half-plane of $\Omega$, seemingly violating causality and Kramers-Kronig relations. Since the integral must instead be performed over a line in the complex plane of $\Omega$ that lies above all of the poles, causality is automatically ensured. This discussion can also be compared with the more formal one in Ref.~\cite{Kirzhnits:1989zz}, which however has a similar physical content.

To conclude, let us briefly recap what we learn from the linear-response approach, and what are its limitations. One finds that a stable system can exchange energy with an external field through a resonance with individual modes. Irreversibility arises from the continuous number of degrees of freedom. An external superluminal field can trigger stable collective superluminal oscillations, while a subluminal field generally triggers Landau-damped collective subluminal oscillations. The latter are \textit{not} normal modes, and thus are not captured by the conventional dispersion relation; they correspond to asymptotic damping that originates from the decoherence of a superposition of many individual normal modes. Instability appears when the dispersion relation has solutions with a positive growth rate, essentially the opposite of Landau damping, where now the energy flows from individual modes to the collective excitation. In the next section, we will connect this viewpoint with the initial-value problem which does not require the assumption of response to an external field.

\section{Fast flavor evolution of an initial condition}\label{sec:initial_condition}

We now turn to the more conventional approach to fast flavor evolution, the ``perturbed initial conditions'' approach, and connect its results with the linear-response theory that we have derived in the previous section. Now there is no external field, so the EOM is Eq.~\eqref{eq:eom_transverse}, but the initial conditions are chosen to be slightly perturbed; the field of flavor coherence is initiated with a nonvanishing and nontrivial $\psi_\bv(t=0)=\overline{\psi}_\bv\ll G_\bv$.

\subsection{Initial-value approach}

For given initial conditions, one can obtain the solution at later times (in the linear regime) by decomposition into the normal modes that we have previously identified, following van Kampen's approach in plasma physics \cite{VanKampen:1955wh}. However, following Landau's approach \cite{Landau:1946jc}, one can solve the initial-value problem much more directly by the method of Laplace transform that we recap in Appendix~\ref{sec:Landau} for the plasma case. Without loss of generality, we assume the initial condition to be of the spatial plane-wave form $\psi_\bv(t=0)=\overline{\psi}_\bv\propto e^{i\bK\cdot\br}$ and introduce the Laplace transform of the sought function
\begin{equation}
    \psi^s_\bv=\int_0^{\infty} dt\, \psi_\bv e^{-st},
\end{equation}
where $s$ is a complex variable. Following standard methods, the EOM in Eq.~\eqref{eq:eom_transverse} becomes
\begin{equation}
    (s+i\bK\cdot\bv-i G\cdot v)\psi^s_\bv=\overline{\psi}_\bv-i\psi^s\cdot v\, G_\bv,
\end{equation}
where $\psi^{s,\mu}$ is the Laplace transform of $\psi^\mu=\int d^2\bv\,v^\mu\psi_\bv$. For notational convenience, once more we introduce the four-vector $K=(is,\bK)$ and $k=K+G$, so that
\begin{equation}
    k\cdot v\,\psi^s_\bv=i\overline{\psi}_\bv+\psi^s\cdot v\,G_\bv.
\end{equation}
Notice the similarity to the normal-mode expression Eq.~\eqref{eq:normal_mode_condition}, although now the initial condition appears on the rhs. Moreover, we can now divide by 
$k\cdot v$ with impunity because $k^0=is+G^0$ is intrinsically complex. Multiplying with $v^\mu$ and integrating over $d^2\bv$ yields
\begin{equation}
    \psi^{s,\mu}=A^\mu+\psi^{s,\nu}\chi^{\mu}_\nu
    \quad\hbox{with}\quad
    A^\mu=i\int \frac{\overline{\psi}_\bv v^\mu}{k\cdot v}d^2\bv
    \quad\hbox{and}\quad
     \chi^\mu_\nu=\int \frac{G_\bv v^\mu v_\nu}{k\cdot v}d^2\bv.
\end{equation}
We have used here the same letter $\chi$ as for the flavor response function of Sec.~\ref{sec:linear_response}; we will see that they coincide on the real axis and thus are indeed the same quantity. The solution for $\psi^s$ is now
\begin{equation}
    \psi^{s,\mu}=(\delta^{\mu}_\nu-\chi^{\mu}_\nu)^{-1}A^\nu,
\end{equation}
where $(\ldots)^{-1}$ denotes matrix inversion. Notice the formal similarity with the linear response to an external field. Finally, inverting the Laplace transform, the solution is
\begin{equation}
    \psi^\mu=\int \frac{ds}{2\pi i}\,\psi^{s,\mu} e^{s t},
\end{equation}
where the integral must be performed on a path in the complex plane such that $\mathrm{Re}(s)$ is to the right of any singularity of the integrand and for $-\infty<\mathrm{Im}(s)<+\infty$. Thus, in terms of $\Omega=K^0=is$, we may write
\begin{equation}\label{eq:explicit_solution}
    \psi^\mu=-\int_{-\infty+i\sigma}^{+\infty+i\sigma} \frac{d\Omega}{2\pi}e^{-i\Omega t}(\delta^\mu_\nu-\chi^\mu_\nu)^{-1} A^\nu,
\end{equation}
where $\sigma$ is a positive number 
larger than the real part of any pole of the integrand.

The integrals appearing in the definitions of $A^\mu$ and $\chi^\mu_\nu$ are always well-defined for $\mathrm{Im}(\Omega)>0$, which is where the integral in Eq.~\eqref{eq:explicit_solution} is performed. As $\mathrm{Im}(\Omega)=\epsilon>0$ tends to zero, all such integrals tend to $\int d^2\bv\,f(\bv) /(k\cdot v+i\epsilon)$, where $\epsilon$ is the (positive) infinitely small imaginary part of $\Omega$. This shows that indeed on the real axis the function $\chi^\mu_\nu$ introduced here coincides with the flavor response function introduced in Sec.~\ref{sec:linear_response}. For $\mathrm{Im}(\Omega)\leq 0$, the functions can still be made well-defined by the procedure of analytical continuation. In any integral of the form
\begin{equation}\label{eq:typical_form_integral}
    \int d^2\bv\, \frac{f_\bv}{\omega-\bk\cdot \bv}=\int_{-1}^{-1} dz \int_0^{2\pi} d\phi\, \frac{f(z,\phi)}{\omega-|\bk||\bv| z},
\end{equation}
where $z$ is the cosine of the angle between $\bv$ and $\bk$, and $\phi$ the azimuthal angle around $\bk$, we must always choose the path of integration in the complex plane of $z$ to pass \textit{below} the singularity $z=\omega/|\bk||\bv|$. For $\mathrm{Im}(\omega)>0$ this means integrating along the real axis, but for $\mathrm{Im}(\omega)\leq 0$ the contour must be deformed to pass below the singularity.

After this brief mathematical excursion we are ready to connect to physics and interpret the results in terms of collective behavior. Let us assume at first that there is no instability. For large $t$, we can deform the integration contour in $\Omega$ making $\sigma$ as small as possible, i.e., until we hit a singularity of the integrand function at a value that we call $\Omega_i$, leading to a term in $\psi^\mu$ behaving as $e^{-i\Omega_i t}$. Thus, the singularities of the integrand functions correspond to asymptotic behaviors of the solution. Assuming for the moment that $A^\nu$ and $\chi^\mu_\nu$ have no singularity -- which we will see is not true -- the dominant singularities come from the points where the matrix $\delta^{\mu}_\nu-\chi^\mu_\nu$ is not invertible. Thus, the collective oscillations correspond to 
\begin{equation}\label{eq:dispersion_relation}
    \Phi_{\omega, \bk}=\mathrm{det}\left[g^{\mu\nu}-\chi^{\mu\nu}\right]=0,
\end{equation}
as in the linear-response approach, but we can now give a new and more general meaning to this equation. Earlier, $\omega$ was real -- the frequency of the applied external field -- and if the dispersion equation was satisfied signalled the existence of self-consistent oscillations. In the initial-condition approach, the values of $\omega_i$ satisfying this equation correspond to asymptotic behaviors of the solution which at large $t$ behave as $e^{-i\Omega_i t}$. 

Crucially, for $\mathrm{Im}(\Omega)<0$, Eq.~\eqref{eq:dispersion_relation} does \emph{not} coincide with the often-used Vlasov-style dispersion relation for normal modes stated after Eq.~\eqref{eq:normal_mode_condition} because the integrals over $d^2\bv$ must always pass below the singularity $k\cdot v=0$. In particular, the solutions of Eq.~\eqref{eq:dispersion_relation} do \emph{not} appear in complex conjugate pairs.

If the system is stable, the zeros of the dispersion relation lie in the lower half-plane, with $\mathrm{Im}(\Omega_i)<0$. Hence, they describe collective oscillations that are damped by a mechanism identical with the well-known Landau damping of plasma waves. We stress that these damped modes are not eigenmodes of the system, and would not appear as a solution of Eq.~\eqref{eq:normal_mode_condition}, which is clear already from the fact that Eq.~\eqref{eq:normal_mode_condition} admits complex solutions only in conjugate pairs. Rather, Landau-damped collective oscillations arise from the superposition of many non-collective (Case-van Kampen) modes that oscillate with different real $\Omega$; they quickly decohere to give an exponentially damped evolution.

Finally, let us turn to the case of an unstable distribution. Here the dispersion relation has zeros in the upper half-plane and hence at late times the dominant behavior is exponentially growing. We recover that if there are instabilities, an initial condition will generally evolve at late times according to the growing solution of the dispersion relation Eq.~\eqref{eq:dispersion_relation}, which in the upper half-plane coincides with the conventional Eq.~\eqref{eq:normal_mode_condition}.

Let us now make the conceptual connection between the conventional approach based on the search for normal modes, which led us to the dispersion relation in Eq.~\eqref{eq:normal_mode_condition}, and the initial-value approach which led us to Eq.~\eqref{eq:dispersion_relation}. In passing we note that the latter is of course identical to the dispersion relation for collective modes obtained in linear-response theory, since the evolution is largely independent of whether we trigger it with an initial perturbation or with an external field. 

The normal-mode approach corresponds to searching for solutions with an exponential dependence on time. As long as one is primarily interested in unstable solutions, one assumes that from any initial condition, the fastest-growing mode will quickly dominate. In general, to find the solution one needs to decompose the initial condition in normal modes, but this requires to include both the discrete collective and the continuum of non-collective (Case-van Kampen) normal modes \cite{VanKampen:1955, Capozzi:2019lso}, which are not well appreciated in the fast flavor literature. Collective modes are either real, or appear in pairs of complex conjugate solutions. The non-collective modes form a continuum for any subluminal mode with $\mathrm{Im}(\omega)=0$ and $\omega^2<|\bk|^2$. They are of no direct physical interest because their singular wave functions cannot be individually excited. They represent a mathematical tool in the continuum limit to expand the initial condition, propagate forward in time, and recombine them to study the evolution of a general initial condition.

Therefore, it is Landau's initial-value approach that most directly captures the time evolution. The solutions of Eq.~\eqref{eq:dispersion_relation} reveal the asymptotic behavior at late times. If the system is unstable, and there are growing modes, then both approaches predict that these will dominate the late evolution, although in the initial-value approach, Eq.~\eqref{eq:explicit_solution} offers an explicit expression for the solution at all times in terms of an integral over the initial condition. If the system is stable, the initial-value approach recovers the dominant Landau damping, which in the normal-modes approach is much harder to identify because it arises from a superposition of many Case-van Kampen modes.

Another analogy which may perhaps shed light on Landau damping, and may be more familiar to particle physicists, is the notion of quasi-stationary states in quantum mechanics. If a particle is enclosed in a very high potential well which however goes to zero at infinity -- the typical example would be the $\alpha$ decay of a nucleus as treated by Gamov~\cite{Gamow:1928zz}, with the $\alpha$ particle trapped within the Coulomb barrier -- the eigenstates are rigorously continuous. However, because the potential well is very high, we may speak with good approximation of metastable states, where the particle is confined within the well for a long time. The energy of such states has an imaginary part, describing their decay rate, and in a formal sense they correspond to a superposition of many real-energy eigenstates of the continuous system. In this analogy, the latter correspond to the Case-van Kampen modes, while the metastable states with exponential damping correspond to the Landau-damped configurations.

\subsection{Energy exchange and instability}
\label{sec:energy}

An instability corresponds to the growth of the transverse component of the polarization vectors, namely the growth of $|\psi^\mu|$. Since the transverse component is associated with its own energy, this growth must happen at the expense of some other form of energy. For weakly unstable modes, the energy derives entirely from neutrinos that are resonant with the given modes. To prove this point, we split the energy in three separate pieces
\begin{equation}
    \mathcal{U}=\mathcal{K}+\mathcal{U}_z+\mathcal{U}_\bot,
\end{equation}
with
\begin{equation}
    \mathcal{K}=\int d^3\br d\Pi\,|\bp| n_\bp,\quad
    \mathcal{U}_z=\int d^3\br\,\frac{P^z_\mu P^{z,\mu}}{4},\quad 
    \mathcal{U}_\bot=\int d^3\br\,\frac{\psi^*_\mu \psi^\mu}{4}.
\end{equation}
An instability must correspond to a growing absolute value of the transverse energy $\mathcal{U}_\bot$, namely
\begin{equation}
    \frac{1}{\mathcal{U}_\bot}\frac{d\mathcal{U}_\bot}{dt}>0.
\end{equation}
The EOM reveals the derivative 
\begin{equation}
    \frac{d\mathcal{U}_\bot}{dt}=\frac{1}{4}\int d^3\br (\psi^*_\mu \partial_t \psi^\mu+\psi^\mu \partial_t \psi^*_\mu).
\end{equation}
In the linear regime, we can consider an individual spatial Fourier mode, so that $\psi_\mu\propto e^{i\bK\cdot\br}$. From the EOM, we then find for the transverse energy rate-of-change
\begin{equation}
    \frac{d\mathcal{U}_\bot}{dt}=\frac{i V}{4}(\psi^*_\mu G^\nu \psi_{\nu}^\mu-\psi_\mu G^\nu \psi_{\nu}^{*,\mu}+K_i\psi^{i\mu}\psi^*_\mu-K_i\psi^{*,i\mu}\psi_\mu),
\end{equation}
where we have introduced $\int d^3\br=V$ the (infinite) volume, $\psi^{\mu\nu}=\int d\Pi\,\psi_\bv v^\mu v^\nu$, and the repeated Latin index $i$ implies summation over the spatial parts.

It is straightforward to see that the first two terms cancel exactly with $d\mathcal{U}_z/dt$ and the last two terms cancel exactly with $d\mathcal{K}/dt$, making the total energy conserved~\cite{Fiorillo:2024fnl}. Introducing the four-vector $\tilde{k}=(G^0,\bG+\bK)$, we can write more compactly
\begin{equation}\label{eq:derivative_1}
    \frac{d\mathcal{U}_\bot}{dt}=\frac{iV}{4}(\psi^{*,\mu}\tilde{k}^\nu \psi_{\mu\nu}-\psi^{\mu}\tilde{k}^\nu \psi^*_{\mu\nu}).
\end{equation}
This expression is exact, since it comes entirely from the EOM. If the system has an unstable mode, we can now express
\begin{equation}
    \psi_{\mu\nu}=\int d^2\bv \psi_{\bv}v_\mu v_\nu\simeq \int d^2\bv \frac{G_\bv v_\mu v_\nu v_\alpha \psi^\alpha}{k\cdot v};
\end{equation}
after replacing in Eq.~\eqref{eq:derivative_1}, we find
\begin{equation}
    \frac{d\mathcal{U}_\bot}{dt}=\frac{iV}{4} \overline{k}_\nu \psi_\mu \psi^*_\alpha \int d^2\bv\,  v^\mu v^\nu v^\alpha G_\bv \left[\frac{1}{k\cdot v}-\frac{1}{k^*\cdot v}\right].
\end{equation}
If the mode is weakly damped, by definition the imaginary part of $\omega$ is small, so in $k\cdot v=\mathrm{Re}(\omega)-\bk\cdot\bv+i\mathrm{Im}(\omega)$ we can take the limit $\mathrm{Im}(\omega)\to0$ and approximate
\begin{equation}\label{eq:final_derivative_energy}
    \frac{d\mathcal{U}_\bot}{dt}\simeq\frac{\pi V}{2}\tilde{k}_\nu \psi_\mu \psi^*_\alpha \int d^2\bv\,v^\mu v^\nu v^\alpha G_\bv 
    \delta\bigl[\mathrm{Re}(\omega)-\bk\cdot\bv\bigr].
\end{equation}
Hence, as advertised earlier, the energy gain $d\mathcal{U}_\bot/dt$ for weakly unstable modes comes entirely from neutrinos that are resonant with the mode. In Sec.~\ref{sec:angular_crossing}, this insight will be the key idea behind an intuitive understanding of the relation between angular crossings and instabilities.

\subsection{Analytic properties of the susceptibility}\label{sec:analytic_properties}

Next we turn to comparing the properties of the susceptibility $\chi$ appearing in the two dispersion relations, the normal-modes one Eq.~\eqref{eq:normal_mode_condition}, and the initial-value one Eq.~\eqref{eq:dispersion_relation}. For practical purposes, it will be useful to rewrite these quantities in terms of the (generically complex) phase velocity $u=\omega/|\bk|$ instead of the frequency $\omega$. Therefore, we write the four-dimensional vector $k^\mu=\kappa(u,\bn)$, where $\kappa=|\bk|$ is the module of the wavevector, $\bn$ is the unit vector in the direction of the wave, and the dispersion relation as
\begin{equation}
    \Phi(u,\kappa,\bn)=\mathrm{det}\bigl[\kappa g^{\mu\nu}-\tilde{\chi}^{\mu\nu}\bigr]=0
\end{equation}
with
\begin{equation}
    \tilde{\chi}^{\alpha\beta}=\int d^2\bv \frac{G_\bv v^\alpha v^\beta}{u-\bn\cdot\bv+i\epsilon}.
\end{equation}
In this form, for a fixed value of $u$, the dispersion relation amounts to finding the eigenvalues of the (complex) pseudo-Euclidean tensor $\tilde{\chi}^{\alpha\beta}$. Eigenvalues that are real and positive correspond to a possible  $\kappa$ and thus to a solution of the dispersion relation.

We start by noting that for any component of $\tilde{\chi}^{\alpha\beta}$ we can separate the integral over $d^2\bv$ into an integral over the cosine of polar angle $z=\bn\cdot\bv$ with the direction $\bn$, and the azimuthal angle $\phi$ around this direction. After integrating over the latter, we always have integrals of the form
\begin{equation}\label{eq:definition_chi_tilde}
    \tilde{\chi}^{\alpha\beta}=\int_{-1}^{+1} dz\,\frac{1}{u-z+i\epsilon}\int_0^{2\pi} d\phi\,G_\bv v^\alpha v^\beta=\int_{-1}^{+1} dz\,\frac{F^{\alpha\beta}}{u-z+i\epsilon}.
\end{equation}
The function $F^{\alpha\beta}$ for real values of $z$ vanishes everywhere for $|z|>1$. Therefore, we may extend it for complex values of $z$ such that it also vanishes for any $|\mathrm{Re}(z)|>1$, while it is equal to the analytical extension of $F^{\alpha\beta}$ for $|\mathrm{Re}(z)|<1$. Thus, the function is discontinuous in the complex plane along the lines $\mathrm{Re}(z)=\pm 1$. On the other hand, these discontinuities
do not generally imply a corresponding singularity for the function $\tilde{\chi}^{\alpha\beta}(u)$, so we now study its analytic properties separately.

For $|\mathrm{Re}(u)|<1$, the integrand function in $\tilde{\chi}^{\alpha\beta}(u)$ has a pole at $z=u$. The $i\epsilon$ serves as a reminder that the contour of the $dz$ integration, in principle the real axis, should be deformed so as to pass below the pole $z=u$. If $\mathrm{Im}(u)>0$ this deformation is not necessary and the susceptibility is anyway finite. As $\mathrm{Im}(u)$ becomes infinitesimally close to $0$ and the pole finally touches the real axis, the $i\epsilon$ prescription takes the contour of integration below the pole and the function $\tilde{\chi}^{\alpha\beta}(u)$ remains analytic for any $\mathrm{Im}(u)\geq 0$.

If $\mathrm{Im}(u)<0$, the path of integration must still be chosen to pass below the pole $z=u$; this can always be done if $F^{\alpha\beta}(z)$ does not have any poles in the lower half-plane. If there is such a pole, say at $z=z_i$, then if $u$ comes close to $u=z_i$, the two poles $z=u$ and $z=z_i$ come together, pinching the contour of integration. Hence, the singularities of $F^{\alpha\beta}(z)$ in the lower half-plane lead to analogous singularities for $\tilde{\chi}^{\alpha\beta}(u)$. However, crucially, close to the real axis $\mathrm{Im}(u)=0$ the susceptibility $\tilde{\chi}^{\alpha\beta}(u)$ is an analytic function for $|\mathrm{Re}(u)|<1.$

A much simpler discussion pertains to $|\mathrm{Re}(u)|>1$. Here, the pole for $z=u$ is in a region of the complex plane where $F^{\alpha\beta}(z)=0$. The contour of integration does not even need to be deformed and the function $\tilde{\chi}^{\alpha\beta}(u)$ is analytic.

\begin{figure}[ht]
    \includegraphics[width=0.48\textwidth]{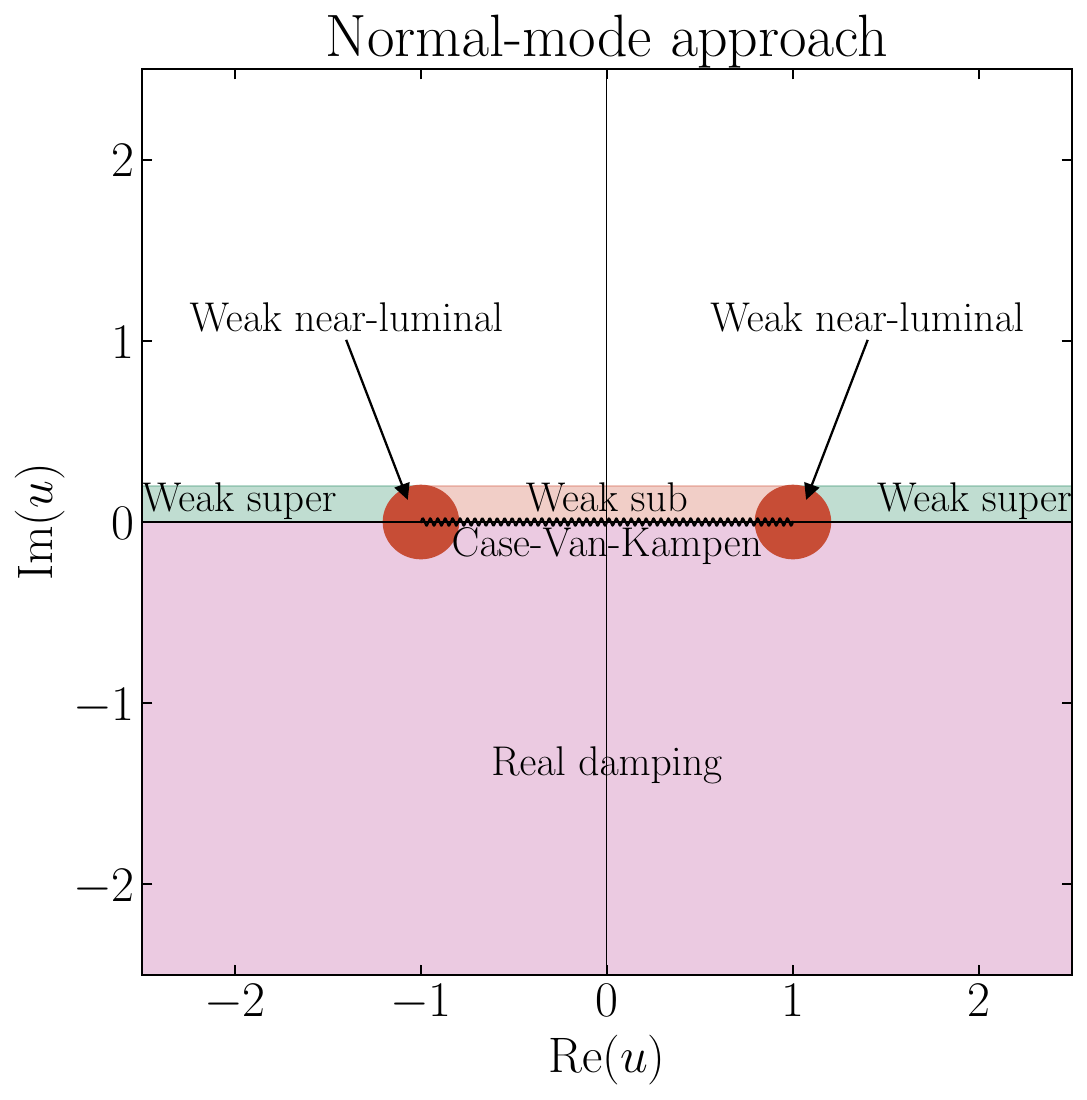}
    \includegraphics[width=0.48\textwidth]{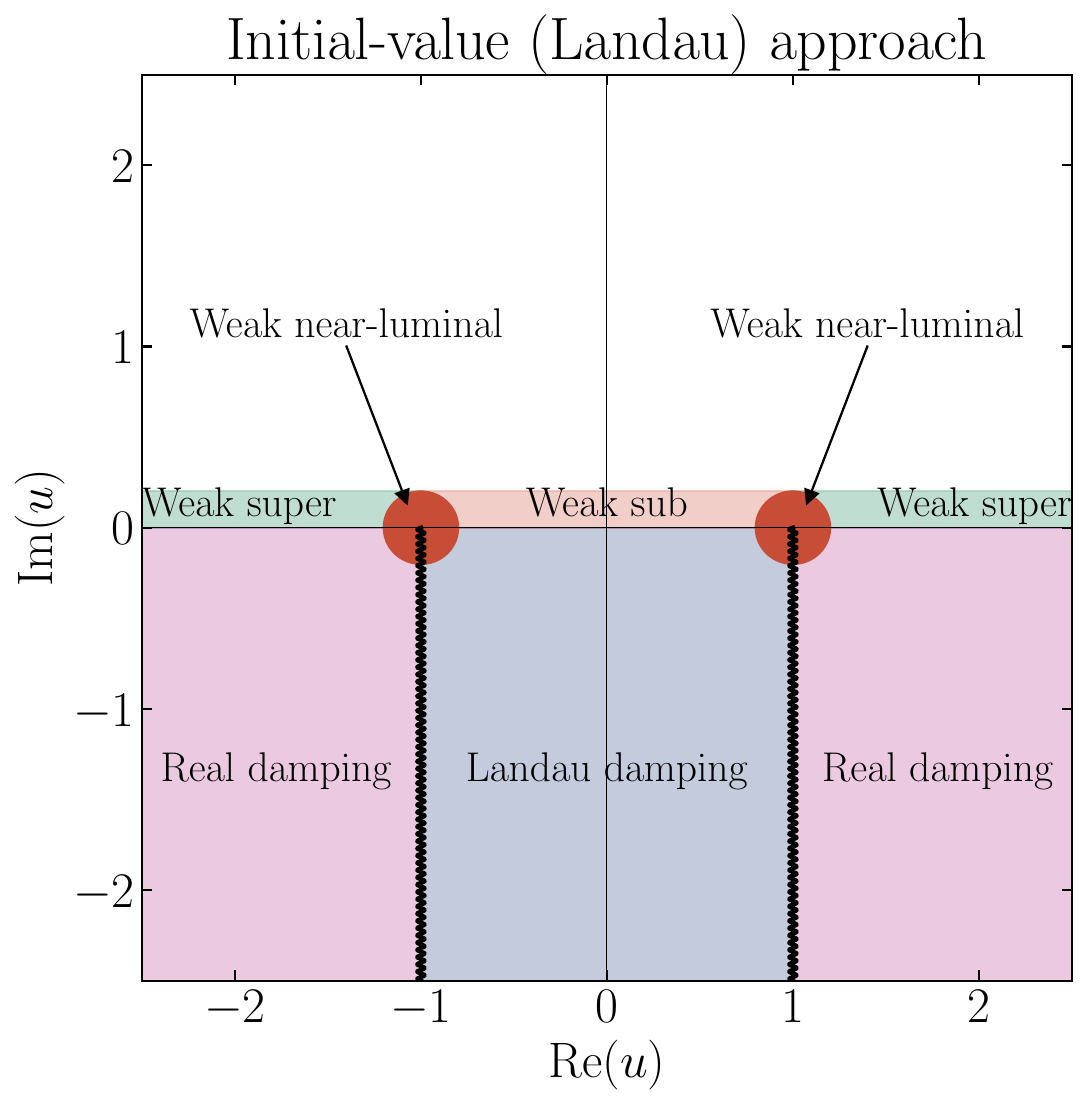}
    \caption{Analytic structure, depending on the complex phase velocity $u=\omega/\kappa$, of the function $\Phi(u,\kappa,\bn)={\rm det}(g^{\mu\nu}-\tilde\chi^{\mu\nu})$ that determines the dispersion relation through $\Phi(u,\kappa,\bn)=0$.\break \textit{Left:} Normal-mode approach including Case-van Kampen modes, leaving a branch cut on the real axis between $-1\leq \textrm{Re}(u)\leq+1$. \textit{Right:} Initial-value (Landau) approach, leaving branch cuts on the vertical lines at $\textrm{Re}(u)=\pm1$ and $\textrm{Im}(u)\leq0$. The red dots highlight the regions of weakly damped, near-luminal modes, where unstable modes are guaranteed for an crossed distribution $G_\bv$.
    }\label{fig:complex_plane}
\end{figure}

A subtle case is $\mathrm{Re}(u)=\pm 1$. If $\mathrm{Im}(u)>0$, the contour of integration need not be deformed and thus the function is still analytic. On the other hand, if $\mathrm{Im}(u)\leq 0$, the pole $z=u$ lies on a singular line of the integrand function $F^{\alpha\beta}(z)$. Thus, $\tilde{\chi}^{\alpha\beta}(u)$ is singular for $\mathrm{Re}(u)=\pm 1$ and $\mathrm{Im}(u)\leq 0$. These lines are branch cuts for the function $\tilde{\chi}^{\alpha\beta}(u)$. We show this situation in the right panel of Fig.~\ref{fig:complex_plane}, where we sketch the analyticity properties of $\Phi(u,\kappa,\bn)$ as a function of the complex phase velocity $u$. There are no poles in the upper half-plane, and the function remains analytic throughout the real axis into the lower half-plane. The vertical branch cuts separate the regions with $|\mathrm{Re}(u)|>1$ (superluminal modes) and $|\mathrm{Re}(u)|<1$ (subluminal modes), where solutions in the lower half-plane are not true eigenmodes, but rather collective Landau-damped motions. 

The left panel shows the corresponding situation for the normal-mode approach, where the dispersion relation is given by Vlasov's expression everywhere and is analytic except on the real axis on the segment $-1\leq \textrm{Re}(u)\leq+1$. The solutions of the dispersion relation are true eigenmodes everywhere except on this segment and, if they are not real, appear with complex conjugate frequencies (or here phase velocities). The segment itself is populated with Case-van Kampen modes and ${\Phi}(u,\kappa,\bn)$ itself formally has a branch cut.

The dispersion relations are identical in the upper half-plane and continuing to the lower half-plane to the left and right of the vertical branch cuts seen in the right panel. In the region between these lines and on and below the corresponding segment of the real axis, the dispersion relations are different and the meaning of the modes is different: (i)~true eigenmodes in the left panel that are truly damped and have complex-conjugate growing counterparts in the upper half-plane
and (ii)~Landau-damped motions that are not eigenmodes of the system.

For our further discussion, the main advantage of the Landau version is that for weakly damped modes with $u$ near the real axis, the function ${\Phi}(u,\kappa,\bn)$ can be expanded as a power series in $u$, except at the exactly luminal points (at the center of the red dots). Weakly unstable modes are the ones where resonances with neutrinos are the driving force, providing an intuitive understanding of why angular crossings guarantee instabilities, although only weak ones. As we will see in our companion paper~\cite{Fiorillo:2024}, the possibility to expand will allow us to obtain accurate approximations for the growth rates of weakly unstable subluminal modes. Still, for modes with $\mathrm{Re}(u)$ very close to $\pm 1$ (near-luminal modes), these expansions are not entirely trivial due to the branch cut being very close.

Now that we have identified the dominant singularities in the Landau dispersion relation, we can be more precise about the asymptotic behavior at late times of a stable system. We previously mentioned that this behavior was dominated by the solutions of the dispersion relation in the lower half-plane of $\omega$ that lie closest to the real axis. These can be either undamped modes (which generally can only be superluminal), or Landau-damped ones. If there are no undamped modes, it would appear that the Landau-damped ones would dominate at late times. However, we have now learnt that there is also a branch-cut singularity in the flavor response function. One can show that in the integral in Eq.~\eqref{eq:explicit_solution} this branch cut leads to a power-law late-time behavior $\psi^\mu \propto t^{-1}$, not exponential suppression. This behavior has a clear physical interpretation; since there it a maximum speed with which neutrinos can move, the ones moving close to the speed of light do not have strong decoherence with one another, and thus lead to a less-than-exponential suppression. Obviously, for unstable systems, for which the dominant late-time behavior is exponentially growing, this does not produce any relevant phenomenological consequence.

\section{Modes close to luminality}

\label{sec:LuminalModes}

Near-luminal modes with $\textrm{Re}(u)\simeq\pm1$ and weak damping, meaning $|\mathrm{Im}(u)|\ll 1$, the modes within the red dots in Fig.~\ref{fig:complex_plane}, play a key role for the understanding of instabilities  because, as soon as an angular crossing is formed, some of these near-luminal modes will grow: they will enter in resonance with neutrinos with polarization vectors ``flipped'' in the region beyond the crossing. However, to understand this effect on a formal level, we will need to know the asymptotic form of the dispersion relation for these near-luminal modes, which is the main topic of the present section.

In the function $\Phi(u,\kappa,\bn)$, the singular behavior is entirely determined by integrals of the form
\begin{equation}\label{eq:example_luminal_integral}
    I=\int d^2\bv\,\frac{f(\bv)}{u-\bv\cdot\bn+i\epsilon}.
\end{equation}
We can always perform the integration over the azimuthal variable around the direction $\bn$ to write it in the form
\begin{equation}
    I=\int_{-1}^{+1} dz\,\frac{F(z)}{u-z+i\epsilon},
\end{equation}
where $F(z)=\int_0^{2\pi} d\phi\,f(z,\phi)$.

Henceforth we will denote $\uR=\mathrm{Re}(u)$ and $\uI=\mathrm{Im}(u)$. For near-luminal modes, we may write $u=1+\delta u\, e^{i\phi}$ with $0<\delta u\ll1$, so that $\uI=\delta u\sin \phi$ and $\uR=1+\delta u \sin \phi$, where $\phi$ is a general phase unrelated to the previous azimuthal angle. A similar expansion can of course be performed close to $u=-1$. The integrals in Eq.~\eqref{eq:example_luminal_integral} are logarithmically divergent as $\delta u\to 0$; we will now show how to extract the leading logarithmic dependence and obtain an approximate expansion for this regime. Actually, to extract the growth rate, it will be necessary to expand to up to orders $\log\delta u$, $\delta u$, and $\delta u \log\delta u$.

We rewrite the integral as
\begin{equation}\label{eq:I-integral}
    I=\int_{-1}^{+1} dz\, \frac{F(z)-F(1)}{u-z+i\epsilon}+F(1)\int_{-1}^{+1} dz\,\frac{1}{u-z+i\epsilon}.
\end{equation}
Notice that $F(1)=2\pi f(\bn)$. The second integral can be done and expanded to first order in $\delta u$ in the form
\begin{equation}
    \int_{-1}^{+1} dz\,\frac{1}{u-z+i\epsilon}
    =\log\left[\frac{2+\delta u\,e^{i\phi}}{\delta u\,e^{i\phi}}\right]
    \simeq\log\left[\frac{2}{\delta u }\right]-i \phi +\frac{\delta u}{2}\, e^{i\phi}.
\end{equation}
In principle, the logarithm requires a specification of branch. However, we can easily see that our choice is the correct one; for $\phi=0$ (slightly superluminal) it produces no imaginary part, while for $\phi=\pi$ (slightly subluminal) it reproduces the correct $i\pi$ prescription from the Landau contour. The branch cut along the line $\mathrm{Im}(u)\leq 0$ and $\mathrm{Re}(u)=1$ is recovered, since $\phi$ jumps from $-\pi/2$ to $3\pi/2$ on the right and left of the branch cut.

In the first integral in Eq.~\eqref{eq:I-integral}, which we call $J$, we can proceed with a further subtraction in the form 
\begin{equation}
    J=\int_{-1}^{+1}\!dz\,\frac{F(z)-F(1)}{u-z+i\epsilon}
    =\int_{-1}^{+1}\!dz\,\frac{F(z)-F(1)}{1-z}
    -\delta u\, e^{i\phi}\int_{-1}^{+1}\!dz\,\frac{F(z)-F(1)}{(1-z)(u-z+i\epsilon)},
\end{equation}
where the first integral converges without $i\epsilon$. The second integral is again logarithmically divergent for $\delta u\to 0$; to capture this divergence, we need to perform a final subtraction so that it becomes
\begin{equation}
   \int_{-1}^{+1}\!dz\, \frac{F(z)-F(1)-F'(1) (z-1)}{(1-z)(u-z+i\epsilon)}+ F'(1)\int_{-1}^{+1}\frac{dz}{u-z+i\epsilon}.
\end{equation}
In the first term, there is no further logarithmic divergence and we can set $u=1$ in the denominator, neglecting terms of higher orders in $\delta u$ as anticipated. The second integral can be done explicitly and expanded as before, so we finally find
\begin{eqnarray}
     J&=&\int_{-1}^{+1}\!dz\,\frac{F(z)-F(1)}{1-z}-\delta u\,e^{i\phi}\int_{-1}^{+1}\!dz\,\frac{F(z)-F(1)-F'(1) (z-1)}{(1-z)^2}
     \nonumber\\[1ex]
     &&{}+\delta u\,e^{i\phi}F'(1)\log\left[\frac{2}{\delta u}\right]-i\delta u\,\phi e^{i\phi}F'(1),
\end{eqnarray}
neglecting terms of order $\delta u^2$.

While we have performed these calculation with a specific parameterization of the integral, we can easily return to a rotation-invariant notation, noting that $F(1)=2\pi f(\bn)$. The derivative takes the value
\begin{equation}
    F'(1)=2\pi\frac{\partial f(\bv)}{\partial (\bn\cdot \bv)}\Big|_{\bv=\bn}.
\end{equation}
With these replacements, arbitrarily close to the luminal sphere, the original integral in Eq.~\eqref{eq:example_luminal_integral} is 
\begin{eqnarray}
    I&=&\int d^2\bv\,\frac{f(\bv)-f(\bn)}{1-\bn\cdot\bv}+2\pi \log\left[\frac{2}{\delta u}\right]\left[f(\bn)+\frac{\partial f(\bv)}{\partial (\bn\cdot \bv)}\Big|_{\bv=\bn}\delta u\, e^{i\phi}\right]
    \nonumber\\[1ex]
    &&{}+\pi f(\bn)\,\delta u\,e^{i\phi}-2\pi i \phi \left[f(\bn)+\frac{\partial f(\bv)}{\partial (\bn\cdot \bv)}\Big|_{\bv=\bn} \delta u\,e^{i\phi}\right]
    \nonumber\\[1ex]
    &&{}-\delta u\,e^{i\phi}\int d^2\bv\,\frac{f(\bv)-f(\bn)+\frac{\partial f(\bv)}{\partial (\bn\cdot \bv)}\big|_{\bv=\bn}(1-\bv\cdot\bn)}{(1-\bv\cdot\bn)^2}.
\end{eqnarray}
While this expression may look much more complicated than the original one, it really is not, since the dependence on $\delta u$ and $\phi$ is completely explicit.

This strategy allows us to approximate the integrals appearing in the dispersion relation, in particular in the tensor $\tilde{\chi}^{\mu\nu}$. 
To this end, we need to replace $f(\bv)\to 2\pi G_\bv v^\mu v^\nu$ and introducing the four-vector $n^\mu=(1,\bn)$ and replacing, we find $f(\bn)\to G_\bn n^\mu n^\nu$, while $F'(1)$ needs to be computed by explicit differentiation, and we define it as $F'(1)=2\pi F^{\mu\nu}$. We will not need the explicit form of this term in the following.

Hence, our key result in this section is the flavor response function near the luminal sphere for $\delta u\to 0$ in the form 
\begin{eqnarray}\label{eq:master_expansion_nearlum}
    \tilde{\chi}^{\mu\nu}&\simeq&\int d^2\bv \frac{G_\bv v^\mu v^\nu-G_\bn n^\mu n^\nu}{1-\bn\cdot\bv}+2\pi\left(\log\left[\frac{2}{\delta u}\right]-i\phi\right)\left(G_\bn n^\mu n^\nu+F^{\mu\nu}\delta u\,e^{i\phi}\right)
    \nonumber\\[2ex]
    &&{}+\pi G_\bn n^\mu n^\nu \delta u\,e^{i\phi} -\delta u\,e^{i\phi}\int d^2\bv\frac{G_\bv v^\mu v^\nu-G_\bn n^\mu n^\nu+F^{\mu\nu}(1-\bv\cdot\bn)}{(1-\bv\cdot\bn)^2}.
\end{eqnarray}
We notice that, as $\delta u\to 0$, even in this near-luminal regime it is still true that the imaginary part of the flavor response tensor is entirely resonant
\begin{equation}
    \mathrm{Im}(\tilde{\chi}^{\mu\nu})\simeq -2\pi \phi G_\bn n^\mu n^\nu
\end{equation}
and thus is dominated by neutrinos moving in the direction of the wave. Hence, the qualitative resonant picture that we have described above is still valid, with neutrinos moving in phase with the wave contributing to the imaginary part of the tensor $\tilde{\chi}^{\mu\nu}$, and thus ultimately to the growth rate. We will now use this result to explore in detail the dispersion relation on the luminal sphere ($\uR=1$) close to an angular crossing.

\section{Angular crossings lead to instabilities}\label{sec:angular_crossing}

Perhaps the only exact result guiding most of the literature on fast flavor conversions is that an angular crossing in the flavor distribution function $G_\bv$ is not only a necessary, but also a sufficient condition for the appearance of instabilities. This result drives most practical studies concerning supernova and neutron-star merger simulations, seeking regions where an angular crossing may lie. Yet, despite being such a foundational result, it can hardly be said that the relevance of angular crossings is deeply understood. The only sufficiency proof~\cite{Morinaga:2021vmc} is mathematically highly involved, and offers little intuitive understanding of the reason for the instability appearance. What we show in this section is that the resonance idea introduced in Sect.~\ref{sec:LuminalModes} offers just such an intuitive understanding. 

\subsection{Connection between crossings and instabilities}

Without pretense at mathematical rigor, we begin with an argument for the connection between crossings and instabilities. The central idea is that energy can be extracted from or given to the transverse motion via the resonant interaction of flavor waves and neutrinos. In Sec.~\ref{sec:energy} we have already computed explicitly the rate of energy transfer from the transverse motion to the longitudinal motion and kinetic energy. In Eq.~\eqref{eq:final_derivative_energy}, we can separate out certain parts that depend on ``global'' properties of the angular distribution, and other parts that depend on ``local'' properties in the region where neutrinos are resonant with the mode. Specifically, $\psi_\mu$ and $G_\mu$ mostly depend on the global properties of the distribution $G_\bv$, integrated over all velocities, whereas under the integral, the dominant part of $G_\bv$ is restricted by the delta function to be evaluated in the resonant region and thus locally. 

This difference is the key to understanding intuitively the connection between the angular crossing and the instability, as well as to understand what its impact. The sign of the energy exchange for a given mode depends on the sign of $G_\bv$ for the neutrinos resonant with that mode, a conclusion that is easily interpreted geometrically since if the polarization vector points downward rather than upwards, the motion of the transverse polarization vector of an individual neutrino is effectively time reversed.

Let us imagine to begin with a distribution that has no angular crossing, but a region where $G_\bv$ is close to~0. Our conservation-law argument implies that the system is stable and so, for any mode, $\mathcal{U}_\bot^{-1}d\mathcal{U}_\bot/dt <0$. Let us now distort $G_\bv$ very slightly so that it changes sign in the region where it was nearly vanishing. We can certainly find a mode that is resonant only with neutrinos in the ``flipped'' region. For such a mode, the global $G_\bv$ properties remain nearly unchanged, whereas, crucially, the sign of $G_\bv$ has \textit{locally} flipped, implying that the mode now extracts, rather than gives, energy from the kinetic and longitudinal part of the energy of the resonant neutrinos. Furthermore, since the corresponding eigenmodes are resonant with these neutrinos, it is the latter that will feel the effect of the instability more strongly, providing an intuitive explanation for the ``removal'' of the angular crossing that was found numerically and theoretically. In the resonant picture, the qualitative explanation is very simple; the modes that are unstable are exactly the ones that resonate with the neutrinos in the flipped region, and thus the latter depolarize completely. 

This heuristic argument does not rigorously prove that an angular crossing is sufficient for an instability. Its main limitation is the reliance on a weak instability, whereas our later proof encompasses the more general case. However, arguably such a physical argument is more informative because it provides intuition as to why the instability arises. Moreover, the picture of an ``adiabatic'' $G_\bv$ distortion is probably quite realistic because in an astrophysical scenario, an unstable configuration does not suddenly appear, but must be slowly driven by external agents~\cite{Fiorillo:2024qbl}, exactly as in our mind experiment. Our physical argument not only tells us that crossings imply instability, we also learn that the modes that become unstable are the ones primarily resonant in the flipped region.

As a next step, we turn this physical argument into a rigorous proof, which requires some degree of mathematical complexity. To this end, we consider modes that are resonant \textit{only} with neutrinos in the flipped region. The simplest choice are those moving with a phase velocity exactly equal to the speed of light. Such luminal modes can only be resonant with neutrinos moving exactly parallel to them. Our earlier physical reasoning suggests that luminal modes pointing into the flipped region will be resonantly enhanced, while luminal modes pointing in the non-flipped region will not.

To prove this idea mathematically, we will need to make use of some properties of the dispersion relation, which we will adopt in the form
\begin{equation}
    \mathrm{det}\bigl[\kappa g^{\mu\nu}-\tilde{\chi}^{\mu\nu}(\uI,\bn)\bigr]=0,
\end{equation}
where we explicitly show that $\tilde{\chi}^{\mu\nu}$ depends on the direction $\bn$ of the mode and on the imaginary part $\uI$ of the phase velocity, whereas the real part is fixed to $\uR=1$. We will be interested only in modes with $0<\uI\ll1$, for which we can extract only the leading logarithmic terms from Eq.~\eqref{eq:master_expansion_nearlum}. We also neglect terms containing products of $G_\bn$ and $\uI$, since close to the crossing line both quantities are small. We separate the flavor response tensor into its real and imaginary part $\tilde{\chi}^{\mu\nu}=a^{\mu\nu}-i b^{\mu\nu}$ with
\begin{subequations}\label{eq:expansion_luminal_proof}
\begin{eqnarray}
   a^{\mu\nu}&=&\int d^2\bv\,\frac{G_\bv-G_\bn}{1-\bn\cdot\bv}\,v^\mu v^\nu
   -2\pi \log(\uI) G_\bn n^\mu n^\nu,
   \\[1ex]
   b^{\mu\nu}&=&\pi^2 G_\bn n^\mu n^\nu+2\pi \uI \log(\uI) F^{\mu\nu}.
\end{eqnarray}    
\end{subequations}
The problem corresponds to finding the real eigenvalues of the symmetric tensor $\tilde{\chi}^{\mu\nu}$ in a pseudo-Euclidean space caused by the Lorentz metric $g^{\mu\nu}$. We first recall some properties of such tensors (see, e.g., Ref.~\cite{Landau:1975pou}). 

\subsection{Symmetric tensors in pseudo-Euclidean space}

For a symmetric tensor $T^{\mu\nu}$, one can show in the usual way that, for two different eigenvalues $\lambda_1$ and $\lambda_2$, the corresponding eigenvectors must be orthogonal. In $T^{\mu\nu} a_{1,\nu}=\lambda_1 a_1^\mu$ and $T^{\mu\nu} a_{2,\nu}=\lambda_2 a_2^\mu$, we may multiply the first equation with $a_{2,\mu}$, the second with $a_{1,\mu}$, and using the symmetry of $T^{\mu\nu}$, after subtracting we easily find $a_{1,\mu} a_2^\mu=0$. However, even if $T_{\mu\nu}$ is real, some of the eigenvalues can be complex, a behavior caused by the pseudo-Euclidean metric. However, since the coefficients of the determinant equation are all real, the eigenvalues are either real or appear in complex conjugate pairs. 

As a consequence, at least two eigenvalues of a real symmetric tensor $T^{\mu\nu}$ must be real. This is seen if we assume that there is a complex conjugate pair of eigenvalues $\lambda$ and $\lambda^*$, implying that the two eigenvectors $n$ and~$n^*$ are orthogonal. Writing $n=a+ib$ one easily infers that for $n^*_\mu n^{\mu}=0$, one of the four-vectors $a$ and $b$ must be time-like and the other space-like. Thus, if we have two complex conjugate eigenvalues, their eigenvectors span a plane composed of a time-like and a space-like direction. Since the other two eigenvectors of the four-dimensional matrix must be orthogonal to this plane, both of them must be space-like, and therefore their corresponding eigenvalues must be real.

There is one other possibility, namely one eigenvector being light-like, corresponding to two roots of the determinant equation being equal. However, even in this case the light-like vector selects a plane containing a time-like and a space-like direction, and the other two eigenvectors must still be orthogonal to it and therefore space-like with real eigenvalues.

\subsection{Luminal modes on two sides of a crossing}

We now return to the physical problem of identifying unstable modes on the luminal sphere, with $\uR=1$. The idea behind the proof sketched earlier is that modes pointing in the flipped region are unstable, motivating us to consider luminal modes pointing very close to the crossing line, i.e., the direction on the unit sphere where $G_\bv=0$. We expect that if they point on opposite sides of this line, they will either be damped or enhanced. 

To prove this behavior, we first show that there always exist real modes ($\uI=0$) pointing to the crossing line, i.e., modes that are neither growing nor Landau-damped. For such a direction $\bn_c$, we have $G_{\bn_c}=0$ and thus, $b^{\mu\nu}$ in Eq.~\eqref{eq:expansion_luminal_proof} vanishes for $\uI=0$; no neutrinos are in resonance with such a mode. The corresponding response tensor
\begin{equation}
    \tilde{\chi}_c^{\mu\nu}=a_c^{\mu\nu}=\int d^2\bv \frac{G_\bv v^\mu v^\nu}{1-\bn\cdot\bv}
\end{equation}
is purely real, and hence admits at least two real eigenvalues. This means that for $u=1$ there are at least two values of $\kappa$ that are eigenvalues; we denote one of them by $\kappa_i$.

What happens if we slightly move from the crossing line in a direction $\bn$ close to~$\bn_c$? We expect that the corresponding eigenvalue will either disappear or will acquire a small imaginary part $u=1+i\uI$, while the value of $\kappa=\kappa_i+\delta\kappa$ will slightly change. Thus, the dispersion relation changes to
\begin{equation}
    \mathrm{det}\bigl[(\kappa_i+\delta\kappa)g^{\mu\nu}-a^{\mu\nu}(\uI,\bn)+ib^{\mu\nu}(\uI,\bn)\bigr]=0.
\end{equation}
For $\bn\simeq \bn_c+\delta \bn$, with $\delta\bn$ small, both $G_\bn$ and $\uI$ are also small and we can expand 
\begin{equation}
    a^{\mu\nu}=a_c^{\mu\nu}+\delta a^{\mu\nu}
    \quad\textrm{and}\quad
    b^{\mu\nu}=n_c^\mu n_c^\nu \pi^2 G_\bn +2\pi \uI \log(\uI) F_c^{\mu\nu} .
\end{equation}
We have evaluated the tensor $F^{\mu\nu}$ directly at the crossing point $F_c^{\mu\nu}$, since it appears multiplied by $\uI$ which is already a small quantity, so one does not need to keep the small difference $\delta\bn$ between $\bn$ and $\bn_c$. In $G_\bn$ we cannot maintain this simplification, since $G_\bn$ vanishes at the crossing point. We will not need the precise form of $\delta a^{\mu\nu}$, although it is easily found from Eq.~\eqref{eq:expansion_luminal_proof}.

By definition, we have $\mathrm{det}\left[\kappa_i g^{\mu\nu}-a_c^{\mu\nu}\right]=0$. Our perturbed determinant equation is
\begin{equation}
\mathrm{det}\bigl[(\kappa_i g^{\mu\nu}-a_c^{\mu\nu})+(\delta\kappa g^{\mu\nu}-\delta a^{\mu\nu}+ib^{\mu\nu})\bigr]=0.
\end{equation}
In expanding the determinant, we notice that the unperturbed matrix $M_c^{\mu\nu}=\kappa_i g^{\mu\nu}-a_c^{\mu\nu}$ has a vanishing eigenvalue; we denote its corresponding eigenvector by $q_i^\mu$. If we bring $M_c^{\mu\nu}$ to its diagonal form, one can see that the linear expansion of the determinant corresponds to taking the product of all the non-vanishing eigenvalues $\lambda$ of $M_c^{\mu\nu}$ times the projection of the perturbation $\delta\kappa g^{\mu\nu}-\delta a^{\mu\nu}+ib^{\mu\nu}$ over the eigenvector $q_i^{\mu}$
\begin{equation}
    q_i^\mu q_i^\nu(\delta\kappa g_{\mu\nu}-\delta a_{\mu\nu}+ib_{\mu\nu})\prod_{\lambda\neq 0} \lambda=0.
\end{equation}
The real and imaginary part of this expression must both vanish, imposing two separate conditions for the two unknowns $\delta\kappa$ and $\uI$. We focus on the vanishing of the imaginary part $q_i^\mu q_i^\nu b_{\mu\nu}=0$. 

To advance further, we need to show that at least one of the eigenvectors of $M_c^{\mu\nu}$ has $q_i^\mu n_{c,\mu}\neq 0$. We proceed by cases:
\begin{itemize}
    \item If all eigenvalues of $M_c^{\mu\nu}$ are real, they correspond to four orthogonal eigenvectors, and so at least two of them are not orthogonal to the light-like vector $n_c^\mu$.
    \item If one of the eigenvectors $q_\mu$ of $M_c^{\mu\nu}$ is light-like, all of the eigenvectors can be orthogonal to $n_c^\mu$ only if $q_\mu= n_{c,\mu}$, but this requires the matrix $M_c^{\mu\nu}$ to satisfy the four conditions $M_c^{\mu\nu}n_\nu=\lambda n_{c,\mu}$. While this might coincidentally happen on one point on the crossing line, it cannot happen on the entire line, since the single parameter $\lambda$ is not sufficient to accommodate the four conditions.
    \item If two of the eigenvalues of $M_c^{\mu\nu}$ are complex conjugates, we denote by $q_J$ the (space-like) eigenvectors for the remaining two real eigenvalues, with $J=1,2$. If both of these eigenvectors were orthogonal to $n_c^\mu$, they would both have to lie in the plane transverse to the direction $\bn_c$. If we denote by $e_{1}^\mu$ and $e_{2}^\mu$ the two four-vectors spanning this transverse plane, this means that we should have $q_{J}^\mu=\cos\theta_J e_{1}^\mu+\sin\theta_J e_2^\mu$ for some $\theta_J$, so the matrix $M_c^{\mu\nu}$ should satisfy the eight conditions
    \begin{equation}
        M_c^{\mu\nu}(\cos\theta_J e_{1,\nu}+\sin\theta_J e_{2,\nu})=\lambda_J (\cos\theta_J e_1^\mu+\sin\theta_J e_2^\mu),\; J=1,2
    \end{equation}
    for some values of $\theta_1$, $\theta_2$, $\lambda_1$, $\lambda_2$. These four parameters are not sufficient to satisfy these eight conditions, thus showing that we can always find a point on the crossing line such that $q_\mu n_c^\mu \neq 0$.
\end{itemize}
This argument neglects the possibility of special symmetries that may enforce the existence of eigenvectors with $q_\mu n_c^\mu=0$. In our companion paper~\cite{Fiorillo:2024}, in which we focus more on axisymmetric distributions, we will show that for this case one such eigenvector always exists. However, this symmetry can only protect one eigenvector, and only for axisymmetric distributions; for the remaining eigenvectors, the counting argument applies. Barring the special symmetry-protected case, our proof holds and we are assured of the existence of at least one eigenvector such that $q_{i,\mu} n_c^\mu\neq 0$.

For such an eigenvector, the condition $q_{i,\mu} q_{i,\nu} b^{\mu\nu}$ implies
\begin{equation}\label{eq:imaginary_part}
    \uI \log \uI=-\frac{\pi  G_\bn (q_i\cdot n_c)^2}{2 F_{c,\mu\nu}q_i^\mu q_i^\nu}.
\end{equation}
For positive $\uI$ arbitrarily small, this equation admits a solution for $\uI$ only if the rhs is negative. Since $G_\bn$ changes sign  through the crossing line, it follows that on one of the two sides this expression will certainly be negative, proving that an unstable luminal mode exists on one of the two sides. One may worry that the quantity $F_{c,\mu\nu}q_i^\mu q_i^\nu$ could vanish. While this could only happen accidentally, and thus certainly not on the entire crossing line, even in this case one only needs to expand Eq.~\eqref{eq:master_expansion_nearlum} beyond the dominant logarithmic term to see that the conclusion remains unchanged. The fundamental point is that in the imaginary part $b^{\mu\nu}$ the only term that does not vanish as $\uI\to 0$ is $\pi^2 G_\bn n_c^\mu n_c^\nu$, which changes sign through the crossing line, so the remaining part of $b_{\mu\nu}$ depending on $\uI$ must change sign as well to make $b_{\mu\nu}q_i^\mu q_i^\nu=0$.

We have proven that the crossing lines are edges for regions of instability, meaning that in passing through a crossing line on the luminal sphere on one side we will find unstable modes. It is also easy to see that these lines are the only possible edges on the luminal sphere; if $\uI\to 0$ without $G_\bn$ also vanishing, the dispersion relation can never be satisfied, due to the infinite logarithmic terms. Thus, on the luminal sphere, crossing lines are the only ones that can separate a region of instability from one of stability.

Our line of argument in some sense parallels Morinaga's proof~\cite{Morinaga:2021vmc}, which also considered luminal modes pointing close to the crossing line. We should therefore explain what are the differences:
\begin{itemize}
    \item We use a dispersion relation that is analytic, allowing us to use the tools of analysis (the non-analytic components are entirely encoded in the logarithmic terms that we have found explicitly).  Hence, while Morinaga used certain algebraic properties of the coefficients of polynomials to prove the existence of a complex solution, we derive explicitly the imaginary part in the $\uI\to 0$ limit. Our proof is thus in principle constructive, although from the practical point of view it may be quite complicated to explicitly perform such a calculation.
    \item We provide an intuitive reason why angular crossings form instabilities. As seen from Eq.~\eqref{eq:imaginary_part}, the imaginary part $\uI$ grows with $G_\bn$, the distribution function of the neutrinos moving in resonance with the waves. Modes moving on opposite sides of the crossing are resonant with neutrinos with opposite signs of the distribution, explaining why only on one side of the crossing there can be an instability.
    \item We have proven not only that the crossing lines are edges for unstable regions, but also that on the luminal sphere they are the only possible edges. This allows us to infer more properties on the region of unstable wavenumbers for a given angular distribution. In our companion paper~\cite{Fiorillo:2024}, we will use this result to deduce that axisymmetric, single-crossed angular distributions always have some unstable modes directed along the axis of symmetry.  
\end{itemize}

\section{Discussion}\label{sec:discussion}

The aim of this work was mainly to reconsider the foundations of neutrino fast collective flavor evolution. In the past, the problem of understanding this effect has been mainly tackled by either analyzing small deviations from equilibrium with linear stability analysis, or by numerically solving toy systems. While these approaches are well suited to obtain concrete answers to practical problems, they can be less ideal for building intuition for the origin of the instability. Here, we show that many features can be grasped intuitively, without the need to resort to highly involved mathematical proofs or numerical solutions, both of which can have the character of black boxes.

Contrary to most previous works, we start with systems that are stable. The fast evolution here manifests itself in the existence of collective flavor waves. Superluminal collective waves may remain stable, whereas subluminal collective waves are generally exponentially Landau damped. The physical origin of this effect is neutrinos resonantly drawing energy from the collective wave; neutrinos with different velocities keep oscillating, but with rapidly decohering phases, explaining the damping. In addition, the presence of a maximal velocity -- the speed of light -- implies a special form of power-law Landau damping associated with luminal modes.

Stability itself, as usual in theoretical physics, is enforced by conservation laws. The fast-flavor system is generally not in a state of minimum energy, but it is in a state of maximum angular momentum if it has no crossing. This insight provides an intuitive and immediate proof of the necessity of crossings for the instability.

An instability thus requires a crossing of the angular distribution, and one may imagine to begin with an uncrossed spectrum and slowly distort it until it develops a region where it flips sign. When this happens, the transverse polarization vector of a neutrino in the flipped region behaves effectively in the time-reversed way from before, thus growing, the opposite of Landau damping. Hence, we come to understand instabilities as a resonant transfer of energy to the transverse motion of neutrinos in the flipped region. These resonant neutrinos are the ones that deviate more from the original flavor-diagonal configuration, so one expects the largest effect exactly in this region, providing an intuitive link with the removal of the angular crossing numerically observed and theoretically predicted. This is of course a purely intuitive argument, since our approach is purely linear and cannot describe the evolution once the perturbation becomes non-linear; in that regime, the only proof is our quasi-linear approach~\cite{Fiorillo:2024qbl}. Our resonant picture is applied only to the case of fast instabilities, and we do not deal with the question of whether a similar framework could be helpful also in the case of slow instabilities.

We have also shown that the collective behavior of the system is understood in terms of its linear response to an external field. In fact, this is the most physical way of thinking about the initiation of the instability; we have shown that the off-diagonal components of the mass term, rapidly varying in the basis which accounts for the refractive matter effect, acts as an external flavor field triggering the instability. In this sense, the usual viewpoint of perturbed initial conditions is artificial, although in a practical sense the instability probably evolves without depending much on the nature of the seeding. Nevertheless, the external-field perspective allows for some conceptual advantages, since we can describe the properties of the neutrino gas in terms of a flavor response function, and consequently a flavor susceptibility. 

Collective oscillations arise as poles of the flavor susceptibility. This is the point where the analogy with plasma physics is most instructive. Tools similar to those arising in plasma physics were also adapted in Ref.~\cite{Capozzi:2017gqd} to fast conversions, but the focus was on the difference between absolute and convective instabilities, and the dispersion relation was still entirely rooted in the normal-mode approach, without considering the modified dispersion relation with well-defined analytical properties and leading to Landau damping.

Our theoretical framework, based on the two pillars of conservation laws and resonant energy transfer, allows us to build intuition for the origin and evolution of fast flavor instabilities. However, if one is not satisfied with intuition and seeks rigor, we also use our approach to obtain a new proof that an angular crossing \textit{necessarily} leads to an instability. Compared to Morinaga's earlier proof, we do not rely on algebraic properties of polynomials, but rather on the tools of analysis. Moreover, it is constructive in the sense of allowing in principle for an explicit determination of the growth rates of the modes close to instability. We show that the crossing lines on the unit sphere of neutrino velocities form an edge for unstable modes; luminal modes can be unstable only on one side of the crossing line. This matches of course with the intuitive proof based on resonant energy transfer.

Overall, to obtain a physics-informed recipe for the effect of fast flavor conversions, numerical solutions are certainly of great help, but perhaps cannot conclusively be generalized to realistic astrophysical systems over hydrodynamical scales. A theoretical understanding of the evolution remains a key requirement. We believe that the two key ideas introduced here -- conservation laws and resonance -- may form the basis of such an understanding. In our companion paper~\cite{Fiorillo:2024}, we will show how the resonant behavior may be used to directly obtain approximate expressions for the growth rates of weakly unstable distributions, presumably the only ones actually arising in truly self-consistent astrophysical systems.

\acknowledgments

We thank Basudeb Dasgupta, Luke Johns, Hiroki Nagakura, Shashank Shalgar, G\"unter Sigl, Irene Tamborra, Meng-Ru Wu, and Zewei Xiong for important comments on the manuscript. DFGF is supported by the Villum Fonden under Project No.\ 29388 and the European Union's Horizon 2020 Research and Innovation Program under the Marie Sk{\l}odowska-Curie Grant Agreement No.\ 847523 ``INTERACTIONS.'' GGR acknowledges partial support by the German Research Foundation (DFG) through the Collaborative Research Centre ``Neutrinos and Dark Matter in Astro- and Particle Physics (NDM),'' Grant SFB-1258-283604770, and under Germany’s Excellence Strategy through the Cluster of Excellence ORIGINS EXC-2094-390783311.

\appendix

\section{Plasma waves for pedestrians}

\label{sec:Landau}

\subsection{Longitudinal plasma waves (Langmuir waves)}

Our study in the framework of linear-response theory is inspired by the similarity of flavor waves with other forms of collective behavior that arise from a combination of transport by particle flow with a coherent interaction among the streaming particles. The classic case is the one of longitudinal plasma waves (Langmuir waves) in a gas of electrons on a static homogeneous background of positively charged ions. We use this traditional example to go through the same steps as in the main text as a case study that may be both more familiar and more intuitive. For a pedagogical treatment of these topics, see, e.g., Ref.~\cite{Sagan:1993es}.

The starting point is the collisionless Vlasov equation \cite{Vlasov:1945} for the electron velocity distribution $F(t,\br,\bv)$, defined such that $F(t,\br,\bv)d^3\bv$ is the number density of particles in the velocity interval $d^3\bv$
\begin{equation}
    \partial_t F+\dot\br\cdot\bpartial_\br F+\dot\bv\cdot\bpartial_\bv F=0.
\end{equation}
This equation has the character of a continuity equation, where $\bv=\dot\br$ is the electron velocity and $\dot\bv$ can be interpreted as a force causing a drift in momentum space. Notably, this is spawned by the electric field ${\bf E}(t,\br)$ that exists if the electron distribution is disturbed from its homogeneous equilibrium distribution. In this way, disturbances of $F(t,\br,\bv)$ act back on themselves, leading to plasma oscillations or instabilities.

We may write the distribution as $F(t,\br,\bv)=n_e \left[f_0(\bv)+f(t,\br,\bv)\right]$ in terms of the electron density $n_e$, where $f_0(\bv)$ is the undisturbed velocity distribution normalized to $\int f_0(\bv) d^3\bv=1$ and $f(t,\br,\bv)$ a small disturbance. To linear order, one finds
\begin{equation}\label{eq:perturbed_vlasov_equation}
    \partial_t f+\bv\cdot\bpartial_\br f=-\frac{e\,n_e}{m_e}\,{\bf E}[f]\cdot\bpartial_\bv f_0,
\end{equation}
where the electric field is a Coulomb integral over the disturbance $f$ that acts as a source. Without interactions, the rhs vanishes and the equation describes freely streaming electrons in the form of a continuity equation.

We seek normal modes for $f(t,\br,\bv)$ of the form $f(\bv)\,e^{-i(\omega t-\bk\cdot\br)}$, where $f(\bv)$ also depends on $\{\omega,\bk\}$ and we use the same letter $f$ for the disturbance in coordinate space and the amplitude of the normal mode. The EOM becomes
\begin{equation}\label{eq:EOM3}
    (\omega-\bv\cdot\bk)\, f(\bv)=-\omega_{\rm P}^2\,\frac{\bk\cdot\bpartial_\bv f_0(\bv)}{\bk^2}
    \int d^3\bv'\, f(\bv'),
\end{equation}
where $\omega_{\rm P}^2=4\pi\alpha n_e/m_e$ is the square of the nonrelativistic plasma frequency, and we have used Maxwell's equations to write the electric field as
\begin{equation}
    {\bf E}[f]=e\frac{\bk}{\bk^2}\int d^3\bv' f(\bv').
\end{equation}

This equation shows the general structure of the problems we are studying, i.e., the Vlasov term on the lhs and an integral over the velocity distribution of the disturbance on the rhs.

For a given solution of real $\{\omega,\bk\}$, there exist some electrons for which the lhs vanishes, meaning that their velocity component in the direction of $\bk$ matches the phase velocity $u=\omega/\kappa$ of the plane-wave disturbance and are in this sense resonant with the wave. This resonance is at the heart of our discussion. In a plasma, it allows for the transfer of energy between individual electrons and the wave, which is a correlated or coherent motion of all electrons. This effect can lead to damping of the wave (Landau damping), but also to its exponential growth (two-stream instability), depending on the velocity distribution $f_0(\bv)$. The collective motion depends on the 3D velocity distribution of the particles that support the wave and thus has the same general structure that we have found for flavor waves. 

\subsection{Vlasov's dispersion relation}

Ignoring at first such resonances, we can read from Eq.~\eqref{eq:EOM3} the form of the solution to~be
\begin{equation}
     f(\bv)=a\,\frac{\bk\cdot\bpartial_\bv f_0(\bv)}{\bk^2\,(\omega-\bv\cdot\bk)},
\end{equation}
where $a$ is a global factor. Notice that the integral in Eq.~\eqref{eq:EOM3} is a number that does not depend on $\bv$. Inserting this form of the solution on both sides of Eq.~\eqref{eq:EOM3} reveals the self-consistency condition, first derived in 1945 by Vlasov \cite{Vlasov:1945},
\begin{equation}\label{eq:plasmadispersion}
    1=-\frac{\omega_{\rm P}^2}{\bk^2}\int d^3\bv\,\frac{\bk\cdot\bpartial_\bv f_0(\bv)}{\omega-\bv\cdot\bk}.
\end{equation}
Solving this equation for $\{\omega,\bk\}$ provides the dispersion relation $\omega(\bk)$ for plasma waves. If one interprets the integral as a principal value as suggested by Vlasov, one finds the textbook result
\begin{equation}\label{eq:Langmuir-dispersion}
    \omega^2=\omega_{\rm P}^2+\langle v^2\rangle\bk^2,
\end{equation}
where $\langle v^2\rangle=3T/m_e\ll 1$ refers to the electrons. Therefore, plasma waves oscillate with an almost fixed frequency, nearly independently of their $\bk$. For flavor waves, the structure of Eq.~\eqref{eq:plasmadispersion} is analogous, except for a more complicated form of the integral over the velocity distribution of the supporting medium.

The bugbear of Eq.~\eqref{eq:plasmadispersion} consists of the poles that arise from electrons on resonance with the wave. If we only seek exponentially growing modes, where $\omega$ has an imaginary part, the integral is well behaved without problems of interpretation. Moreover, the complex conjugate $\omega^*$ is another solution, so there will be an exponentially growing and a damped solution. One remaining question is the required form of $f_0(\bv)$ that would lead to such instabilities. In the plasma case, one celebrated case is the two-stream instability that arises, for example, when the velocity distribution, projected on one direction $z$,  $\bar f_0(v_z)$, has a bump on the thermal Maxwellian tail (e.g. Ref.~\cite{thorne2017modern}). The exact necessary and sufficient condition for this case was derived by Oliver~Penrose \cite{Penrose:1960}. In the more complicated case of fast flavor waves, the condition for instability is a crossing of the angular lepton-number flux distribution as we have seen in the main text.

Without additional arguments, a Vlasov-type dispersion relation of the form Eq.~\eqref{eq:plasmadispersion} is mathematically meaningful only for unstable modes and for stable (purely real) modes that are superluminal and thus without divergence.\footnote{In the traditional nonrelativistic treatment with a thermal Maxwellian electron distribution, there is no limiting velocity and the integration formally extends to infinity. In this situation, all plasma waves are ``subluminal'' as there are always some electrons on resonance, although in an exponentially suppressed tail of the distribution. This issue arises from the inconsistency of a purely nonrelativistic formulation. However, despite the nonrelativistic treatment, we may imagine that $f_0(\bv)$ is a relativistic version that ends at $|\bv|=1$ and still produces the correct nonrelativistic part.} Historically, Vlasov assumed that the integral should be interpreted in the principal-value sense. The main interest was on stable plasma oscillations, whereas unstable modes such as the two-stream instability were studied only later. 

\subsection{Case-van Kampen modes}

What to make of Vlasov's dispersion relation depends on the problem that one wants to solve. One can identify instabilities, historically the main task in the context of flavor waves. However, one could not solve an initial-value problem, i.e., determine how a given disturbance would evolve in time. To this end one needs a complete set of linearly independent eigenfunctions that would allow one to express a given initial condition. Around ten years after Vlasov's paper, in 1955 this question was resolved by van Kampen~\cite{VanKampen:1955} who showed that besides the collective normal modes that appear as solutions of Eq.~\eqref{eq:EOM3}, there is a continuum that are today known as Case-van Kampen modes and, in the context of fast flavor waves, were called non-collective modes~\cite{Capozzi:2019lso}. Their nature is most easily understood if we consider Eq.~\eqref{eq:EOM3} in the non-interacting limit of a vanishing plasma frequency and thus vanishing rhs, implying $f(\bv)=0$ except for $\omega-\bv\cdot\bk=0$. However, singular solutions exist of the form $f(\bv)=\delta(\omega-\bv\cdot\bk)$. Physically it means that a perturbation with wave vector $\bk$ drifts along and shows a time variation $\omega$ determined by those electrons that fulfill $\omega-\bv\cdot\bk=0$. There is no fixed dispersion law, for any $\bk$ there is a subluminal solution with $-|\bf k|\leq\omega\leq+|\bf k|$. 

For nonvanishing $\omega_{\rm P}$, these modes persist, but with a modified functional form that is still singular,
\begin{equation}
    f(\bv)=\eta(\omega)\,\frac{\cal P}{\omega-\bv\cdot\bk}
    +\lambda(\omega)\,\delta(\omega-\bv\cdot\bk).
\end{equation}
The first term is understood such that under an integral, the Cauchy principal part is taken to interpret the singularity. As shown by van Kampen~\cite{VanKampen:1955} and Case~\cite{Case:1959}, after determining the functions $\eta(\omega)$ and $\lambda(\omega)$, together with the true collective stable or unstable modes, one has found a complete set of linearly independent basis functions.

Any initial perturbation that may have been set up on the plasma can be expanded in these modes and then propagated forward in time with the $e^{-i(\omega t-\bk.\br)}$ factor. If the spectrum includes unstable normal modes, these will dominate after a short while, and this has been the main interest in the context of flavor waves for which an analogous decomposition was performed \cite{Capozzi:2019lso}. If all modes are stable, as for plasma waves in a thermal medium, the superluminal Langmuir wave will persist if present in the original spectrum, whereas the rest of the initial perturbation will dissipate by different Case-van Kampen modes decohering.

The Case-van Kampen modes are less mysterious than they may seem at first. In numerical examples, the distribution function is represented by $N$ discrete beams and the eigenvalue equation amounts to finding the roots and eigenvectors of an $N\times N$ matrix. One then finds discretized versions of the Case-van Kampen modes which are, of course, not singular. In the absence of interactions, one finds $N$ real frequencies $\omega_i$ for given $\bk$, while after including interactions, some collective modes can emerge from the original set, depending on $\bk$ and the interaction strength here expressed by $\omega_{\rm P}$. Some pairs of real modes can merge to form a pair of complex conjugate $\omega$ and $\omega^*$, representing an exponentially growing and damping mode \cite{Capozzi:2019lso}. The singular structure of the Case-van Kampen modes is a mathematical consequence of the continuum limit of the distribution function.

\subsection{Subluminal Langmuir waves}

For longitudinal plasma waves (Langmuir waves) in an isotropic thermal electron gas, one finds the dispersion relation of Eq.~\eqref{eq:Langmuir-dispersion}. For $\omega>|\bk|$, the modes are superluminal and thus undamped. On the other hand, for $\omega<|\bk|$, Vlasov's expression diverges and he assumed that it should be solved by a principal-value interpretation. In this case one finds Eq.~\eqref{eq:Langmuir-dispersion} for both sub- and superluminal Langmuir waves.

However, the subluminal ones are {\em not\/} eigenmodes of the system. Physically, they lose energy to individual electrons (Landau damping). This effect can also be interpreted as Cherenkov absorption of longitudinal electromagnetic waves by electrons that move faster than the wave. In this case, the plasma waves are exponentially damped without a corresponding growing mode. Such modes do not follow from Vlasov's dispersion relation unless one makes the unfounded principal-value assumption, which however does not reveal the damping rate. For weakly damped modes, with a phase velocity much larger than a thermal electron velocity, one may neglect damping and in this sense find both sub- and superluminal Langmuir waves. However, a subluminal Langmuir wave is fundamentally a superposition of Case-van Kampen modes, which are the true eigenmodes, leading to its dissipation by decoherence, which is another picture of Landau damping. Of course, in the original strictly nonrelativistic treatment with no limiting electron velocity, all Langmuir waves would be Landau damped and Vlasov's expression never strictly applied.

\subsection{Linear-response approach}

To obtain the correct prescription to deal with the singularity in Vlasov's dispersion relation Eq.~\eqref{eq:plasmadispersion}, we can follow the linear-response approach, where a disturbance of the plasma is understood as the response to an applied external electric field. If the plasma is stable, we can imagine that the field is monochromatic $\mathbf{E}(t,\br)=\bf E\,e^{-i(\omega t -\bk\cdot \br)+\epsilon t}$; the $\epsilon$ term ensures that the field is adiabatically inserted from the infinite past, enforcing causality. In this case, using Eq.~\eqref{eq:perturbed_vlasov_equation}, the electron distribution function is perturbed by the electric field
\begin{equation}
    f(\bv)=\frac{i e n_e}{m_e(\omega-\bk\cdot\bv+i\epsilon)}\,{\bf E}\cdot\bpartial_\bv f_0,
\end{equation}
where again we write $f(t,\br,\bv)$ as $f(\bv)\,e^{-i(\omega t-\bk\cdot \br)+\epsilon t}$. The time-dependent electron perturbation induces a current 
\begin{equation}
    \mathbf{j}=\int d^3\bv\, e f(\bv)\,\bv;
\end{equation}
in the language of conventional electrodynamics of a medium this response corresponds to a time-dependent polarization ${\bf j}=-i\omega \bP$. In turn, the electric displacement in the medium is ${\bD}={\bf E}+{\bf P}$. Introducing the permittivity tensor $\varepsilon_{ij}$ we write ${D}_i=\varepsilon_{ij}{E}_j$, with spatial indices denoted by Latin letters, and find
\begin{equation}\label{eq:permittivity}
    \varepsilon_{ij}=\delta_{ij}+\chi_{ij}=\delta_{ij}+\frac{\omega_{\rm P}^2}{\omega}\int d^3\bv\,\frac{v_i \partial_{v_j}f_0}{\omega-\bk\cdot\bv+i\epsilon},
\end{equation}
where the $i\epsilon$ prescription is not to be confused with the permittivity~$\varepsilon$. Equation~\eqref{eq:permittivity} connects the collective plasma oscillations with the standard language of linear-response theory in electromagnetism (see Ref.~\cite{Kirzhnits:1989zz} for a review). The susceptibility $\chi_{ij}$, measuring the response of the medium to the field, is exactly analogous to the flavor response function introduced in the main text. As an aside, we should also notice that a relativistic treatment of this problem, while potentially very elegant, is impossible because of our assumption of an electric field alone. A spacetime-varying electric field necessarily induces a magnetic field, which is neglected as usual in the non-relativistic regime, so our kinetic equation is by definition not relativistically invariant. On the other hand, we are here aiming for the simplest case of Langmuir waves, where the electric field is parallel to the wave vector, and thus do not have magnetic fields.

\begin{figure}[ht]
    \centering
    \includegraphics[width=0.3\textwidth]{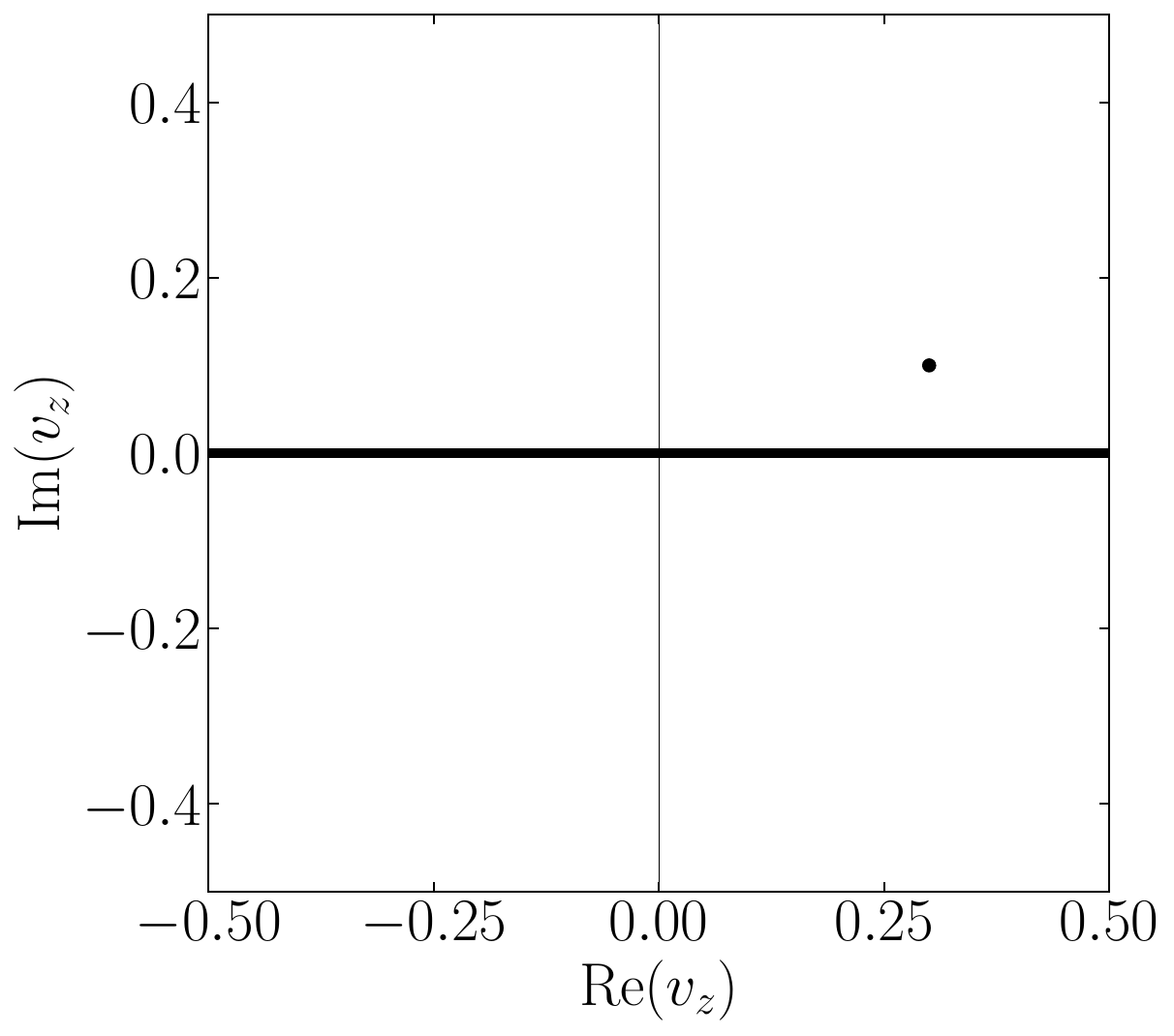}\includegraphics[width=0.3\textwidth]{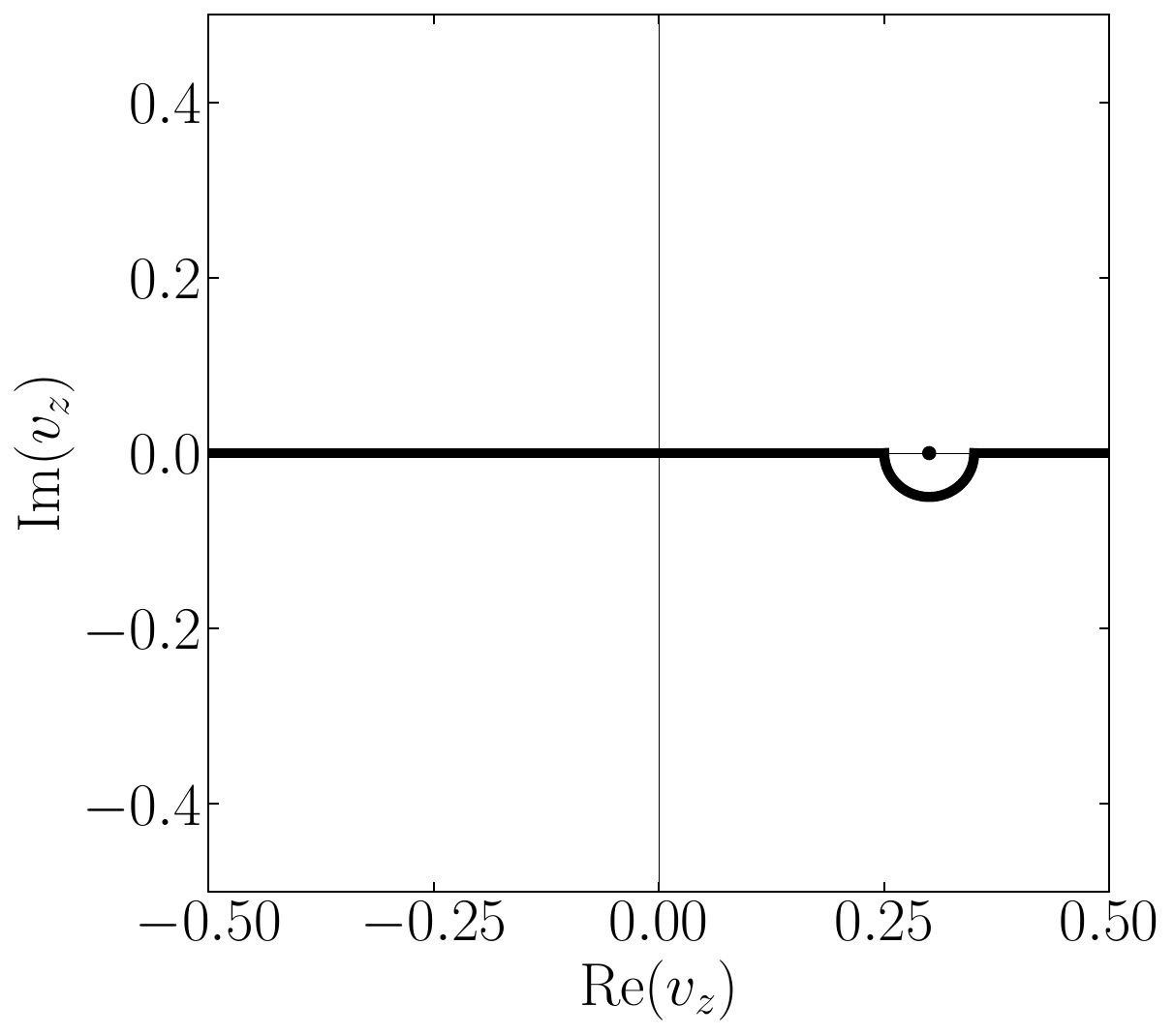}\includegraphics[width=0.3\textwidth]{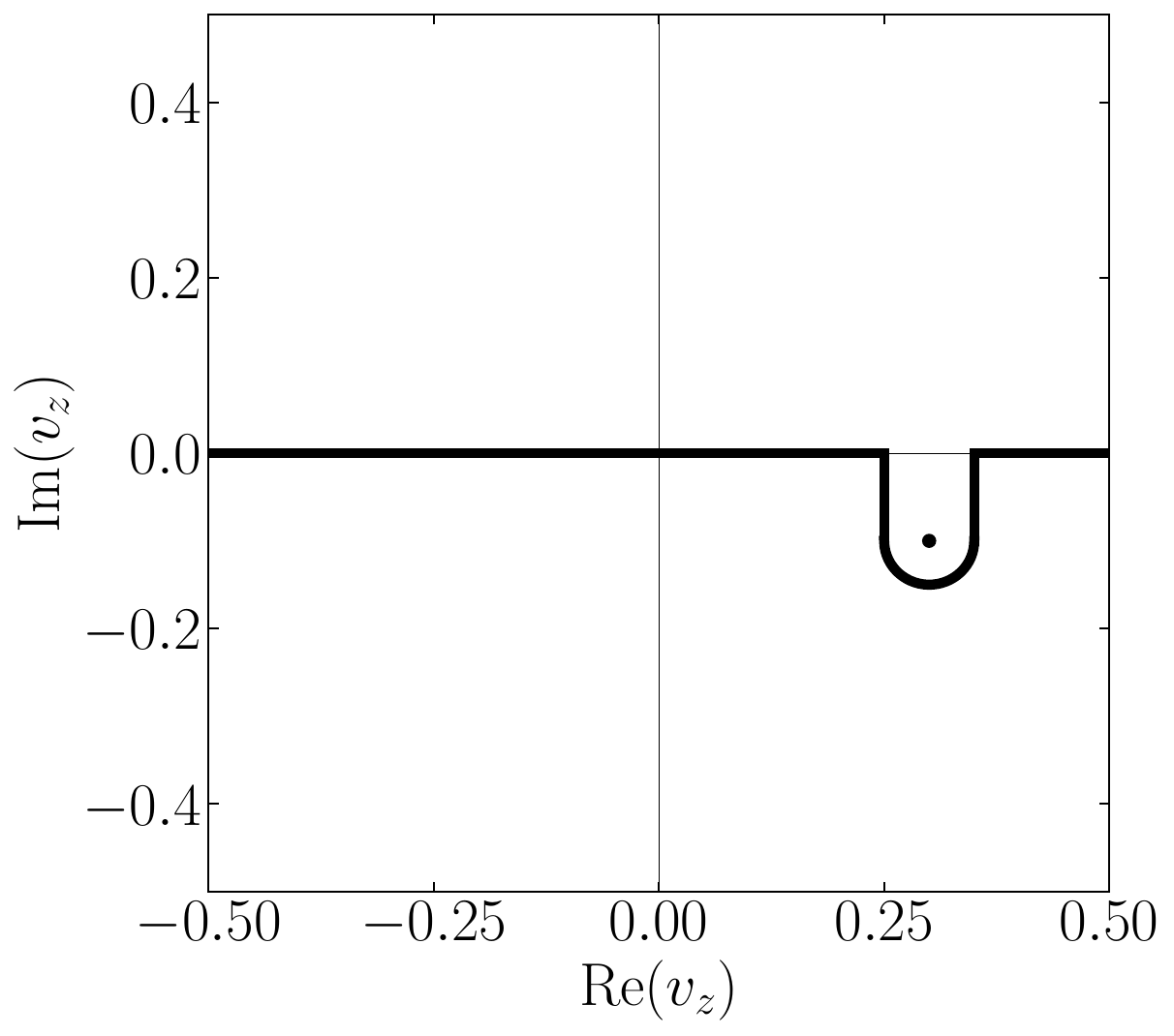}
    \caption{Schematic choice of the integration contours in the plane of the variable $v_z$ for positive, vanishing, and negative (left, center, right) imaginary part $\mathrm{Im}(\omega)$. The contour must always pass below the pole in the integrand.}
    \label{fig:integration_contours}
\end{figure}

The collective modes of the system can now be found by the condition $\mathrm{det}(\varepsilon_{ij})=0$. If $f_0(\bv)$ is isotropic, then the modes can obviously be decoupled into  two degenerate modes transverse to $\bk$, which we do not consider here, and one longitudinal mode polarized along $\bk$. Projecting its dispersion relation along the polarization of the wave, we find $k_i k_j \varepsilon_{ij}=0$, which leads to
\begin{equation}\label{eq:plasmadispersionieps}
    1=-\frac{\omega_{\rm P}^2}{\bk^2}\int d^3\bv\,\frac{\bk\cdot\bpartial_\bv f_0(\bv)}{\omega-\bv\cdot\bk+i\epsilon},
\end{equation}
identical with Vlasov's form of Eq.~\eqref{eq:plasmadispersion} except for the prescription in the denominator of how to integrate around the pole. While the external field is applied with a real frequency, we can now extend the validity of our dispersion relation to identify collective modes with a negative imaginary part for the frequency, corresponding to damping of the wave. We can rewrite Eq.~\eqref{eq:plasmadispersionieps} as
\begin{equation}\label{eq:plasmadispersionieps2}
    1=-\frac{\omega_{\rm P}^2}{\bk^2}\int dv_z\,\frac{1}{\omega-|\bk| v_z+i\epsilon}\int dv_x dv_y\,\bk\cdot\bpartial_\bv f_0(\bv),
\end{equation}
separating the integral over the velocity along the direction of $\bk$, which we call $v_z$, and the other two components. For $\mathrm{Im}(\omega)>0$, the $dv_z$ integral must be performed over the real axis, but the $i\epsilon$ prescription allows us to extend the solution also to $\mathrm{Im}(\omega)\leq 0$, as shown in Fig.~\ref{fig:integration_contours}. As $\mathrm{Im}(\omega)\to 0$, since $\epsilon$ is positive, the contour of integration must be deformed so as to pass below the pole at $v_z=\omega/|\bk|$. Hence, as $\omega$ develops a negative imaginary part, the contour must still be deformed passing below the pole. This prescription, which we obtained naturally by the adiabatic insertion argument, is designed to ensure causality. In fact, since the contour must only be deformed when the integrand has a pole in the lower half-plane, it follows that the only singularities of the \textit{integral} must lie in the lower half-plane, satisfying the Kramers-Kronig assumptions.

We can now use this modified dispersion relation to obtain the rate of Landau damping of the subluminal Langmuir waves introduced above. Using a Maxwellian distribution for $f_0(\bv)$ in Eq.~\eqref{eq:plasmadispersionieps}, and expanding the real part of the integral for $|v_z \bk|\ll \omega$, we recover
\begin{equation}
    1=\frac{\omega_{\rm P}^2+\langle v^2\rangle \bk^2 \omega_{\rm P}^2/\omega^2}{\omega^2}
    -i\,\sqrt{\frac{\pi}{2}}\,  
    \frac{\omega_{\rm P}^2 \omega }{|\bk|^3}
    \left(\frac{m_e}{T}\right)^{3/2}e^{-\frac{m_e \omega^2}{2T \bk^2}}.
\end{equation}
The imaginary part of this expression is exponentially suppressed and therefore to lowest order the dispersion relation is $\omega^2\simeq \omega_{\rm P}+\langle v^2\rangle\bk^2$, the usual dispersion relation for Langmuir waves that one would have obtained from Vlasov's prescription using the pincipal-part prescription of the divergent integral. Solving iteratively we immediately also find the imaginary part
\begin{equation}\label{eq:LandauDamping1}
    \omega^2\simeq \omega_{\rm P}^2\left[1+\frac{\langle v^2\rangle \bk^2}{\omega_{\rm P}^2} 
    -i\,\sqrt{\frac{\pi}{2}}\,  
    \frac{\omega_{\rm P}^3}{|\bk|^3}
    \left(\frac{m_e}{T}\right)^{3/2}
    \,e^{-\frac{m_e (\omega_{\rm P}^2+\langle v^2\rangle \bk^2)}{2T |\bk|^2}}\right].
\end{equation}
This shows that subluminal Langmuir waves are damped, despite not being eigenmodes of the system but rather superpositions of many Case-van Kampen modes. The damping rate is half of the imaginary part of Eq.~\eqref{eq:LandauDamping1} and with $\langle v^2\rangle=3T/m_e$ is
\begin{equation}
    \gamma=\sqrt{\pi}\,e^{-3/2}\,
    \omega_{\rm P} \left(\frac{3\omega_{\rm P}^2}{2\langle v^2\rangle \bk^2}\right)^{3/2}
     \,e^{-\frac{3\omega_{\rm P}^2}{2\langle v^2\rangle|\bk|^2}}.
\end{equation}
When the phase velocity $u\simeq\omega_{\rm P}/|\bk|$ is comparable to or smaller than a typical electron velocity $\langle v^2\rangle^{1/2}$, the exponential factor is no longer small and the real and imaginary part of the eigenfrequency $\omega$ are both of the order of $\omega_{\rm P}$. Formally, this damping rate is largest for $\omega_{\rm P}^2/\bk^2=\langle v^2\rangle$, but as the entire expression was derived in the limit $\gamma\ll\omega_{\rm P}$, it is no longer self-consistent in the strongly-damped regime.

\subsection{Initial-value problem and Landau damping}

The idea that collective plasma oscillations can be damped in the collisionless limit of the kinetic equation goes back to Landau's celebrated 1946 paper \cite{Landau:1946jc}, sometimes called the most important paper in plasma physics ever, that was written in response to Vlasov's 1945 paper, which Landau calls ``mostly incorrect.'' He started from a different premise than our derivation and posed an initial-value problem, asking for the time evolution of a disturbance of the electron distribution that is initially prescribed. (As we have seen, using Case-van Kampen modes, such an initial-value problem can also be solved along the lines of Vlasov, but van Kampen wrote his paper only in 1955 \cite{VanKampen:1955}.)

Following Landau, we consider a single spatial Fourier mode of $f(t,\br,\bv)$ with wavenumber $\bk$ and use the notation $f(t,\bv)\,e^{i\bk\cdot\br}$ so that the EOM is
\begin{equation}\label{eq:EOM4}
    (i\partial_t-\bv\cdot\bk)\, f(\bv)=-\omega_{\rm P}^2\,\frac{\bk\cdot\bpartial_\bv f_0(\bv)}{\bk^2}
    \int d^3\bv'\, f(\bv').
\end{equation}
Landau used the strategy of performing a Laplace transform in time
\begin{equation}
    f_{s}(\bv)=\int_0^{\infty}dt\, f(t,\bv)\,e^{-st},
\end{equation}
where $s$ is a complex variable; if the system is stable this integral converges for any ${\rm Re}(s)>0$. The solution can then be recovered with an inverse Laplace transform
\begin{equation}
    f(t,\bv)=\frac{1}{2\pi i}\int ds\, f_{s}(\bv)\,e^{st},
\end{equation}
where the $ds$ integral must be performed over a path in the complex plane from $s=-i\infty+\sigma$ to $s=i\infty+\sigma$, where $\sigma>0$ is chosen so that the path lies to the right of any singularity of $f_s$. We can now obtain the EOM for $f_s(\bv)$, after using
\begin{equation}
    \int_0^{\infty}\!dt\,\frac{\partial f(t,\bv)}{\partial t}\,e^{-st}=sf_s(\bv)-f(t=0,\bv)
\end{equation}
which is obtained after integration by parts. Hence, the EOM becomes
\begin{equation}
    (s+i\bk\cdot\bv)f_s(\bv)-i\omega_{\rm P}^2 \frac{\bk\cdot\bpartial_\bv f_0(\bv)}{\bk^2}\int d^3\bv'f_s(\bv')=f(t=0,\bv).
\end{equation}
The integral $I_s=\int d^3\bv'\,f_s(\bv')$ is obtained from here to be
\begin{equation}
    I_s=\frac{\int \frac{f(t=0,\bv)}{s+i\bk\cdot\bv}\,d^3\bv}{1-i\int \frac{\omega_P^2}{s+i\bk\cdot\bv}\frac{(\bk\cdot\bpartial_\bv)f_0(\bv)}{\bk^2}\,d^3\bv},
\end{equation}
after which the full solution is
\begin{equation}\label{eq:plasma_full_solution}
   f_s(\bv)=\frac{f(t=0,\bv)}{s+i\bk\cdot\bv}+i\frac{\omega_{\rm P}^2 I_s}{s+i\bk\cdot\bv}\frac{(\bk\cdot\bpartial_\bv)f_0(\bv)}{\bk^2}.
\end{equation}
We can now perform the inverse Laplace transform to obtain the time dependence of the solution. Let us first consider the time dependence of 
\begin{equation}
    I(t)=\int d^3\bv\,{f}(\bv)=\int \frac{ds}{2\pi i}\,e^{st}I_s,
\end{equation}
which is proportional to the total charge density perturbation and therefore to the electric field of the wave. The asymptotic behavior of the integral at late times is dominated by the singularities of the integrand function that lie closest to the axis $\mathrm{Re}(s)=0$; assuming that $f(t=0,\bv)$ and $f_0(\bv)$ are entire functions of their complex arguments, such singularities correspond to the zeros of the denominator. If we call $s=-i\omega$, these zeros coincide with the solutions of the dispersion relation Eq.~\eqref{eq:plasmadispersionieps}. Thus, we recover the result that the collective modes correspond to the asymptotic evolution of the system at late times for a given initial condition. On the other hand, from Eq.~\eqref{eq:plasma_full_solution} we see that the first term, when subject to the inverse Laplace transform, produces an evolution that is not damped but simply oscillates in time as $f(t,\bv)\propto e^{-i\bk\cdot\bv t}$. Thus, while the collective variables (e.g. the electric field) evolve in time according to the solution of the dispersion relation, the single-velocity variables also have undamped evolution corresponding to the Case-van Kampen modes. There is no paradox in this conclusion, since the superposition of the many different Case-van Kampen modes is of course what gives rise to the damped collective motion.

\subsection{Forward-scattering perspective}

Landau damping of plasma waves appears at first surprising because the starting point is a \textit{collisionless} kinetic equation. There is an alternative viewpoint to obtain the dispersion relation for longitudinal plasma waves which is perhaps more familiar to a particle physics community, based on the field theory of the interacting electrons and electromagnetic waves, and leads to the interpretation of Landau damping as Cherenkov absorption. Transverse electromagnetic waves, the usual photons, suffer refraction in a plasma, but such that their resulting four-momentum is superluminal ($\omega^2>\bk^2$) as if they had an effective mass. They are not damped to lowest order in the fine-structure constant. At higher order, they are damped by Compton scattering $\gamma+e\to e+\gamma$, inverse bremsstrahlung, or other processes. However, in a non-plasma medium such as water or air, the interaction with the atomic electrons can render the dispersion relation subluminal ($\omega^2<\bk^2$, refractive index $n>1$ with $n=|\bk|/\omega$). In this case, Cherenkov absorption or emission \hbox{$\gamma+e\leftrightarrow e$} becomes kinematically possible, i.e., a freely propagating electron can emit or absorb photons with subluminal phase velocity. In a plasma, in addition to the usual photons, longitudinal waves exist as collective motions of the electrons. These waves can also be seen as quantized excitations (longitudinal photons), in a nonrelativistic plasma with the dispersion relation $\omega^2\simeq\omega_{\rm P}^2+\langle v^2\rangle\bk^2$. For $\bk^2\gtrsim\omega_{\rm P}^2$, the dispersion relation turns subluminal and the Cherenkov effect allows for the absorption by freely propagating electrons.\footnote{The physical interpretation of Landau damping as Cherenkov radiation of longitudinal plasma excitations is common knowledge today, but we were unable to track down where this idea might have first appeared in print. The original treatment of plasma oscillations, both originally in the Soviet Union with Vlasov's \cite{Vlasov:1945} and Landau's \cite{Landau:1946jc} seminal papers, and a few years later in the West mainly with Bohm and Pines's treatment~\cite{pines1952collective}, was based on the kinetic equation for the electrons. The physical interpretation of Landau damping as the resonant extraction of energy from individual particles was given in Ref.~\cite{bohm1949theory}, but without identifying the Cherenkov process as an interpretation. Probably this realization appeared gradually in the Eastern literature, and it seems that not even the authors working at the time could pinpoint a precise paper putting forward this idea (see the footnote on page 65 in Ref.~\cite{ginzburg1970propagation}). In the first explicit reference that we could track down~\cite{shafranov1958propagation}, the Cherenkov interpretation was treated as generally known.}

\begin{figure}[ht]
\centering
    \includegraphics[width=0.7\textwidth]{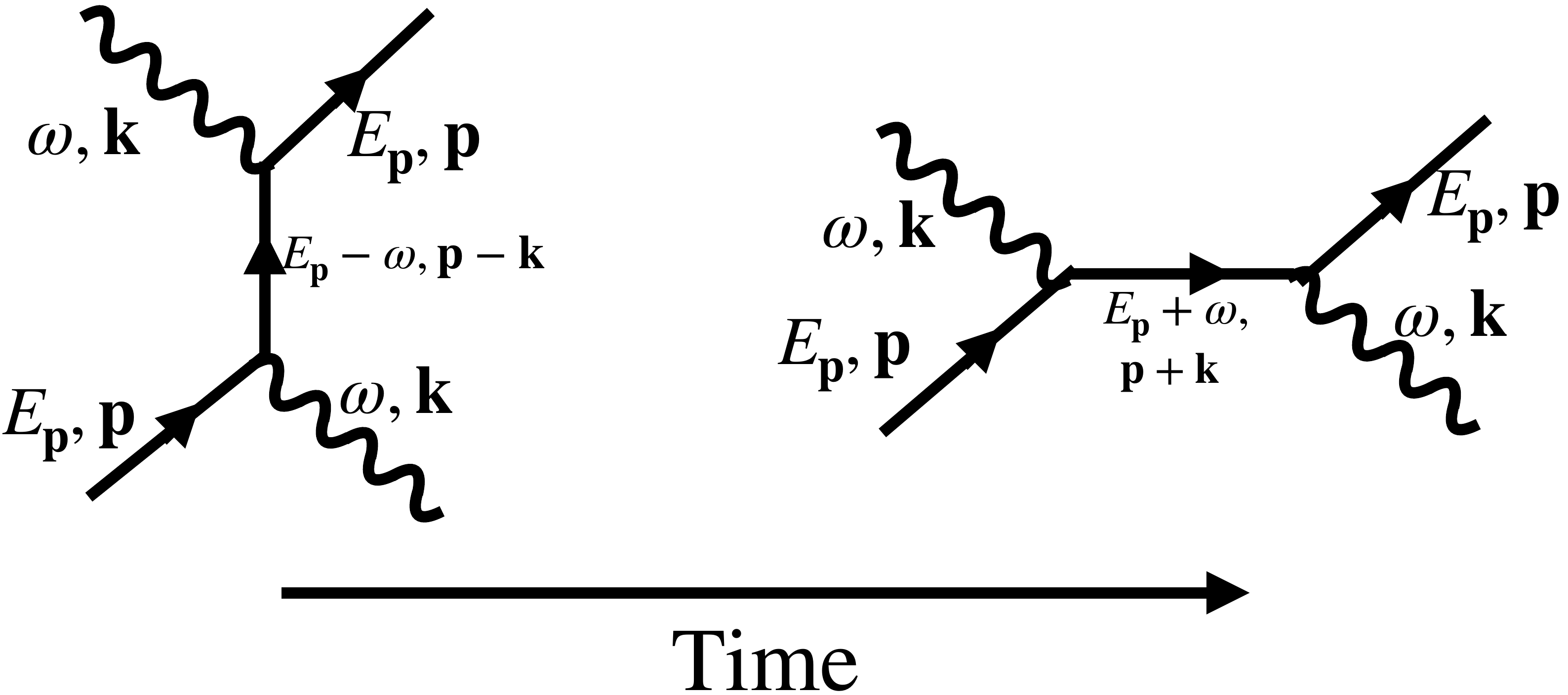}
    \caption{Feynman diagrams for the refractive scattering of longitudinal photons on electrons in the plasma in analogy to the forward Compton scattering of usual transverse photons.
    }\label{fig:scattering}
\end{figure}

The general dispersion relation of plasma waves from the perspective of photon refraction (photon forward scattering on the plasma constituents) can be done elegantly in a relativistically invariant way~\cite{Altherr:1992mf, Braaten:1993jw}, but it is unnecessarily cumbersome to go through the four-dimensional structure for our limited purpose. To see how the dispersion relation of longitudinal plasma waves arises from this picture, it is enough to consider the non-relativistic Lagrangian density
\begin{equation}
    \mathcal{L}=-\frac{|\bNabla \phi|^2}{2}+\psi^\dagger\left[i\partial_t+\frac{\bNabla^2}{2m_e}-e\phi\right]\psi,
\end{equation}
where $\phi$ is the scalar potential (we work in Coulomb gauge $\bNabla\cdot \bA=0$ and neglect the transverse waves associated with $\bA$) and $\psi$ is the electron field. Notice the absence of a kinetic term for $\phi$; longitudinal waves cannot propagate in vacuum. We can now obtain the renormalized dispersion relation for $\phi$ in a medium by folding in the forward scattering amplitude over the electrons; the latter is obtained by computing the diagrams in Fig.~\ref{fig:scattering}, where the energy and momentum of each field is specified. Using the Feynman rules for this theory, after renaming the electron momentum in the second diagram $\bp\to\bp+\bk$, we find that the scattering amplitude summed over all electrons with momentum distribution $f_\bp$, such that $f_\bp d^3\bp$ is the number of particles in the momentum interval $d^3\bp$ (a factor $2$ due to spin is automatically included in the definition of $f_\bp$), is
\begin{equation}\label{eq:amplitude}
    \mathcal{A}=e^2\int d^3\bp \left[\frac{f_\bp-f_{\bp+\bk}}{\omega-\bv\cdot\bk+i\epsilon\,\mathrm{sign}(\omega)}\right]\simeq -e^2\int d^3\bp\,
    \frac{\bk\cdot\bpartial_\bp f_\bp}{\omega-\bv \cdot \bk+i\epsilon\,\mathrm{sign}(\omega)}.
\end{equation}
The denominator here comes from the electron propagator, and is regularized following Feynman's prescription that positive (negative) frequencies propagate forward (backward) in time. The forward-scattering amplitude therefore provides us directly the self-energy for the time-ordered propagator of the longitudinal photon. If we are interested in the retarded self-energy, we can simply replace $i\epsilon\,\mathrm{sign}(\omega)\to i\epsilon$. Hence, in the following, we use this slightly modified prescription, which makes no difference for positive-frequency excitations.
We have also approximated the difference $f_\bp-f_{\bp+\bk}$ with the derivative of the distribution function since the wavelength of the electromagnetic field is assumed much larger than the electron de Broglie wavelength. The forward-scattering amplitude renormalizes the dispersion relation; from the equations of motion for the field $\phi$ we obtain\begin{equation}
    |\bk|^2+\alpha\int d^3\bp\,\frac{\bk\cdot\bpartial_\bp f_\bp }{\omega-\bv \cdot \bk+i\epsilon}=0.
\end{equation}
After calling $f_\bp d^3\bp=n_e f_0(\bv) d^3\bv$, and $\bpartial_\bp=m_e^{-1} \bpartial_\bv$, we easily recover the dispersion relation of Eq.~\eqref{eq:plasmadispersionieps} found in the plasma treatment.

The imaginary part in the dispersion relation arises from the $i\epsilon$ prescription. As well-known, unitarity allows one to relate the imaginary part of the amplitude $\mathcal{A}$ with the rate of scattering of the longitudinal photons on electrons, essentially cutting the diagrams in Fig.~\ref{fig:scattering} along the intermediate line. The corresponding microscopic process is thus Cherenkov absorption or emission of photons, and indeed from Eq.~\eqref{eq:amplitude} we easily see that
\begin{equation}
    \mathrm{Im}(\mathcal{A})=-\pi e^2\int d^3\bp (f_\bp-f_{\bp+\bk})\delta(\omega-\bv\cdot\bk),
\end{equation}
where the difference $f_\bp-f_{\bp+\bk}$ corresponds to the difference between the rate of Cherenkov emission and absorption, and the $\delta$-function enforces conservation of energy in the process.

The identical results between the kinetic and the particle treatment may appear almost fortuitous at first, given the diversity of approach. However, the agreement appears less mysterious once we realize that the scattering amplitude of a field corresponds directly to the additional field generated by the perturbed electrons. Thus, in our kinetic treatment, by computing the perturbation to the electron distribution induced by the field $\phi$, we directly obtained also the additional field produced by such a perturbation, which coincides with the scattering amplitude. This is indeed the same argument by which one can compute the scattering cross section of a classical electromagnetic wave either by determining the electromagnetic field generated by the electrons oscillating within the wave, or by using the Feynman diagrams for a photon scattering off an electron. 

This equivalence suggests that one could in turn obtain, for the case of fast flavor conversion, the dispersion relation from the Feynman diagram for the forward scattering of flavor waves off neutrinos. This framework could be constructed by explicitly separating the flavor waves as an independent degree of freedom, but the results would be identical to the more conventional theory of fast flavor evolution outlined in the main text.

\section{Lepton number conservation from the equations of motion}\label{app:lepton_number_conservation}

The main argument for the necessity of an angular crossing is that the polarization vectors obey the same EOMs as classical spins and that total angular momentum along the $z$-direction in flavor space is conserved, corresponding to the conservation of lepton number. Therefore, if all such spins begin aligned in this direction, the starting point is one of maximal angular momentum and there can be no deviation of any spin without lowering the conserved total. In other words, to have any room for flavor conversion, some of the spins must begin anti-aligned relative to the $z$-direction so that subsequently, the orientation of the different spins can evolve relative to the initial configuration, corresponding to flavor being shuffled around phase space. This symmetry argument, first advanced by Johns in the fast flavor context \cite{Johns:2024bob},\footnote{However, in contrast to the thrust of Ref.~\cite{Johns:2024bob} and in contrast to what is explicitly stated in their abstract, this symmetry argument is unrelated to the idea of ergodicity.}
vastly simplifies earlier dynamical proofs of necessity \cite{Morinaga:2021vmc, Dasgupta:2021gfs} and can be easily extended to slow and collisional instabilities. The far more difficult sufficiency condition is discussed in the main text. What remains here is to perform the formal book keeping explicitly for the necessary condition for the different cases.

The EOMs in the two-flavor case and ignoring the collision term read in the language of polarization vectors
\begin{equation}\label{eq:EOM-Precession}
    v^\alpha\partial_\alpha\vec{P}_\bp
    =\Bigl[\left(\omega_E+v^\alpha\Lambda_\alpha\right)\vec{B}
    +\sqrt{2}\GF v^\alpha\int \frac{d^3\bp'}{(2\pi)^3} v'_\alpha\left(\vec{P}_\bp-\vec{\bar P}_\bp\right)
    \Bigr]
    \times\vec{P}_\bp,
\end{equation}
where $\vec{B}$ is a unit vector in the mass and flavor directions (the $z$-direction) that we have assumed are identical. As usual, the letter $\vec{B}$ suggests a magnetic field around which the flavor spins precess in the absence of collective effects. The refractive effect by normal matter is
\begin{equation}
    \Lambda^\alpha=\sqrt{2}\GF\int \frac{2\,d^3\bp}{(2\pi)^3}
    \Bigl[\left(v_e^\alpha f_{e,\bp}-v_\mu^\alpha f_{\mu,\bp}\right)
    -\left(v_{\bar e}^\alpha f_{\bar e,\bp}-v_{\bar\mu}^\alpha f_{\bar\mu,\bp}\right)\Bigr],
\end{equation}
where the factor of 2 accounts for the two polarization states, we have specifically assumed the $e$ and $\mu$ flavors, and the mass-dependent velocities are $\bv_\ell=\bp/\sqrt{\bp^2+m_\ell^2}$. 

An equation identical to Eq.~\eqref{eq:EOM-Precession} pertains to the antineutrino polarization vectors \smash{$\vec{\bar P}_\bp$} except for a sign change of $\omega_E$. The structure of these equations becomes both more compact and more physically transparent in the flavor isospin convention, where we interpret antiparticles as particles with negative energy and describe their spectrum with negative occupation numbers. In the ultrarelativistic limit, the modes are thus described by $-\infty< E < +\infty$, while their direction of motion $\bv$ remains the physical one defined through $\bp=|E|\bv$ and $v^\alpha = (1,\bv)$. In this way, neutrino phase space is covered only once in a single integration. The modes are represented by $\{E,\bv\}$ with $\vec{P}_{E,\bv}=\vec{P}_\bp$ for $E>0$ whereas $\vec{P}_{E,\bv}=-\vec{\bar P}_{\bp=|E|\bv}$ for $E<0$. The EOM for both neutrinos and antineutrinos then~is
\begin{equation}
    v^\alpha\partial_\alpha\vec{P}_{E,\bv}
    =\Bigl[\left(\omega_E+v^\alpha\Lambda_\alpha\right)\vec{B}
    +\sqrt{2}\GF v^\alpha\int d\Pi' v'_\alpha\vec{P}_{E',\bv'}\Bigr]
    \times\vec{P}_{E,\bv},
\end{equation}
where the phase-space integration is $\int d\Pi=\int_{-\infty}^{+\infty}E^2 dE\int d^2\bv/(2\pi)^3$. The angular integration $\int d^2\bv$ is over the unit sphere. 

Integrating over phase space $\int d\Pi$ on both sides removes the neutrino-neutrino term on the rhs and we are left with
\begin{equation}\label{eq:EOM-Integrated}
    \partial_\alpha  \int d\Pi\,v^\alpha\vec{P}_{E,\bv}
    =\vec{B}\times
    \int d\Pi \left(\omega_E+v^\alpha\Lambda_\alpha\right)\vec{P}_{E,\bv}.
\end{equation}
Taking the scalar product with $\vec{B}$ on both sides puts the rhs to zero and leaves us with
\begin{equation}
    \partial_\alpha \int d\Pi\,v^\alpha P^z_{E,\bv}=0.
\end{equation}
Integrating in addition over all of space provides
\begin{equation}
    \partial_t \int d^3\br \int d\Pi\, P^z_{E,\bv}(t,\br)=-\int d\Pi\, \bv\cdot\int d^3\br\,\bpartial_\br P^z_{E,\bv}(t,\br).
\end{equation}
If there is no net flux through the surface, the rhs vanishes and $P^z=\int d^3\br \int d\Pi\,P^z_{E,\bv}(t,\br)$ is indeed conserved.

We have assumed that initially the system is homogeneous so that $P^z_{E,\bv}(0,\br)$ actually does not depend on $\br$. The quantity
\begin{equation}\label{eq:spectrum}
    G_{E,\bv}=P^z_{E,\bv}(0,0)
\end{equation}
is usually called the spectrum. Therefore, in $\{E,\bv\}$ space it must have positive and negative regions, i.e., it must have zero crossings to allow for any collective motion. The need for spectral crossings thus follows directly from ``angular momentum'' conservation in flavor space.

Morinaga's original proof of necessity \cite{Morinaga:2021vmc} as well as Johns' symmetry argument \cite{Johns:2024bob} were technically formulated only in the fast flavor context and relied on angular momentum conservation (in flavor space), whereas the weaker constraint of angular momentum conservation in the $z$-direction is actually enough and then also encompasses the case of slow conversions driven by the mass term, where the crossing occurs as a function of $E$, a case also covered in Dasgupta's dynamical proof \cite{Dasgupta:2021gfs}, and is of course the original concept of spectral crossing -- see e.g.\ Ref.~\cite{Dasgupta:2009mg}.

Notice, however, that in a strict mathematical sense, both the slow and fast cases require certain abstractions. The slow case relies on the mass and flavor directions being identical (here called the $z$-direction), i.e., the abstraction of vanishing mixing angles
\cite{Airen:2018nvp}. Vacuum mixing angles are known to be large, so this approximation is usually justified by the large matter effect, which however does not make the mixing angles small, but rather causes a fast precession around the flavor direction, or in a corotating frame, a fast precession of $\vec{B}$ around the $z$-axis, essentially nixing, on average, the flavor-off-diagonal piece. It is this picture that motivated us, in the main text, to contemplate the linear response to an external flavor field. The fast flavor case requires the additional abstraction of vanishing mass differences, justified by the idea that fast flavor dynamics is much faster so that the slow dynamics can be ignored on the relevant scales. We are here agnostic as to where in concrete astrophysical environments these idealizations are strictly justified or where they may be questionable -- see e.g.\ Ref.~\cite{Shalgar:2020xns}.

Dasgupta's discussion \cite{Dasgupta:2021gfs} also includes collisions that we have thus far ignored, even though the recently discovered \cite{Johns:2021qby} and now well-established
\cite{Xiong:2022zqz, Liu:2023pjw, Lin:2022dek, Johns:2022yqy, Padilla-Gay:2022wck, Fiorillo:2023ajs} collisional instability also requires a spectral crossing as a function of $E$, similar to slow oscillations. The symmetry argument also applies to this case, although in a slightly more complicated way. The conservation of angular momentum still holds, but in the presence of collisions, the length of the polarization vectors is not conserved, and therefore the state of maximum angular momentum does not necessarily correspond to a unique configuration of spins. However, provided that in the initial state all the species collisionally coupled with the medium are in thermal and chemical equilibrium\footnote{Notice that if all three neutrino flavors are collisionally coupled and in thermal and chemical equilibrium, no instability can develop since the system is already in its equilibrium state. Indeed, in this case no crossing, either angular or spectral, is possible.} -- otherwise the system would trivially evolve, but not due to a flavor instability -- collisions can only damp the transverse components of the polarization vectors, and therefore reduce their total length, as enforced by the second law of thermodynamics. Hence, a state of maximum angular momentum can still only be attained by the polarization vectors all aligned with the $z$-axis, since, if any vector would be tilted, its $z$-component should remain identical, which would amount to having its length increased. Therefore, this argument extends Dasgupta's proof \cite{Dasgupta:2021gfs} to an arbitrary form of the collision term, provided that it conserves $\int d^3\br\,d\Pi\,P^z_\bp$.

\bibliographystyle{JHEP}
\bibliography{References.bib}

\providecommand{\href}[2]{#2}\begingroup\raggedright\begin{thebibliography}{10}

\bibitem{Gribov:1968kq}
V.N.~Gribov and B.~Pontecorvo, \emph{{Neutrino astronomy and lepton charge}}, \href{https://doi.org/10.1016/0370-2693(69)90525-5}{\emph{Phys. Lett. B} {\bfseries 28} (1969) 493}.

\bibitem{Esteban:2020cvm}
I.~Esteban, M.C.~Gonzalez-Garcia, M.~Maltoni, T.~Schwetz and A.~Zhou, \emph{{The fate of hints: updated global analysis of three-flavor neutrino oscillations}}, \href{https://doi.org/10.1007/JHEP09(2020)178}{\emph{JHEP} {\bfseries 09} (2020) 178} [\href{https://arxiv.org/abs/2007.14792}{{\ttfamily 2007.14792}}].

\bibitem{Capozzi:2021fjo}
F.~Capozzi, E.~Di~Valentino, E.~Lisi, A.~Marrone, A.~Melchiorri and A.~Palazzo, \emph{{Unfinished fabric of the three neutrino paradigm}}, \href{https://doi.org/10.1103/PhysRevD.104.083031}{\emph{Phys. Rev. D} {\bfseries 104} (2021) 083031} [\href{https://arxiv.org/abs/2107.00532}{{\ttfamily 2107.00532}}].

\bibitem{deSalas:2020pgw}
P.F.~de~Salas, D.V.~Forero, S.~Gariazzo, P.~Mart\'\i{}nez-Mirav\'e, O.~Mena, C.A.~Ternes, M.~T\'ortola and J.W.F.~Valle, \emph{{2020 global reassessment of the neutrino oscillation picture}}, \href{https://doi.org/10.1007/JHEP02(2021)071}{\emph{JHEP} {\bfseries 02} (2021) 071} [\href{https://arxiv.org/abs/2006.11237}{{\ttfamily 2006.11237}}].

\bibitem{Pantaleone:1992eq}
J.T.~Pantaleone, \emph{{Neutrino oscillations at high densities}}, \href{https://doi.org/10.1016/0370-2693(92)91887-F}{\emph{Phys. Lett. B} {\bfseries 287} (1992) 128}.

\bibitem{Samuel:1993uw}
S.~Samuel, \emph{{Neutrino oscillations in dense neutrino gases}}, \href{https://doi.org/10.1103/PhysRevD.48.1462}{\emph{Phys. Rev. D} {\bfseries 48} (1993) 1462}.

\bibitem{Samuel:1995ri}
S.~Samuel, \emph{{Bimodal coherence in dense selfinteracting neutrino gases}}, \href{https://doi.org/10.1103/PhysRevD.53.5382}{\emph{Phys. Rev. D} {\bfseries 53} (1996) 5382} [\href{https://arxiv.org/abs/hep-ph/9604341}{{\ttfamily hep-ph/9604341}}].

\bibitem{Duan:2009cd}
H.~Duan and J.P.~Kneller, \emph{{Neutrino flavour transformation in supernovae}}, \href{https://doi.org/10.1088/0954-3899/36/11/113201}{\emph{J. Phys. G} {\bfseries 36} (2009) 113201} [\href{https://arxiv.org/abs/0904.0974}{{\ttfamily 0904.0974}}].

\bibitem{Duan:2010bg}
H.~Duan, G.M.~Fuller and Y.-Z.~Qian, \emph{{Collective Neutrino Oscillations}}, \href{https://doi.org/10.1146/annurev.nucl.012809.104524}{\emph{Ann. Rev. Nucl. Part. Sci.} {\bfseries 60} (2010) 569} [\href{https://arxiv.org/abs/1001.2799}{{\ttfamily 1001.2799}}].

\bibitem{Mirizzi:2015eza}
A.~Mirizzi, I.~Tamborra, H.-T.~Janka, N.~Saviano, K.~Scholberg, R.~Bollig, L.~H{\"u}depohl and S.~Chakraborty, \emph{{Supernova Neutrinos: Production, Oscillations and Detection}}, \href{https://doi.org/10.1393/ncr/i2016-10120-8}{\emph{Riv. Nuovo Cim.} {\bfseries 39} (2016) 1} [\href{https://arxiv.org/abs/1508.00785}{{\ttfamily 1508.00785}}].

\bibitem{Tamborra:2020cul}
I.~Tamborra and S.~Shalgar, \emph{{New Developments in Flavor Evolution of a Dense Neutrino Gas}}, \href{https://doi.org/10.1146/annurev-nucl-102920-050505}{\emph{Ann. Rev. Nucl. Part. Sci.} {\bfseries 71} (2021) 165} [\href{https://arxiv.org/abs/2011.01948}{{\ttfamily 2011.01948}}].

\bibitem{Capozzi:2022slf}
F.~Capozzi and N.~Saviano, \emph{{Neutrino Flavor Conversions in High-Density Astrophysical and Cosmological Environments}}, \href{https://doi.org/10.3390/universe8020094}{\emph{Universe} {\bfseries 8} (2022) 94} [\href{https://arxiv.org/abs/2202.02494}{{\ttfamily 2202.02494}}].

\bibitem{Richers:2022zug}
S.~Richers and M.~Sen, \emph{{Fast Flavor Transformations}}, \href{https://doi.org/10.1007/978-981-15-8818-1_125-1}{\emph{Handbook of Nuclear Physics} (2022) 1} [\href{https://arxiv.org/abs/2207.03561}{{\ttfamily 2207.03561}}].

\bibitem{Volpe:2023met}
M.C.~Volpe, \emph{{Neutrinos from dense environments: Flavor mechanisms, theoretical approaches, observations, and new directions}}, \href{https://doi.org/10.1103/RevModPhys.96.025004}{\emph{Rev. Mod. Phys.} {\bfseries 96} (2024) 025004} [\href{https://arxiv.org/abs/2301.11814}{{\ttfamily 2301.11814}}].

\bibitem{Dolgov:1980cq}
A.D.~Dolgov, \emph{{Neutrinos in the early universe}}, {\emph{Sov. J. Nucl. Phys.} {\bfseries 33} (1981) 700}. [{\em Yad.\ Fiz.} {\bf 33} (1981) 1309].

\bibitem{Rudsky}
M.A.~{Rudzsky}, \emph{{Kinetic equations for neutrino spin- and type-oscillations in a medium}}, \href{https://doi.org/10.1007/BF00653658}{\emph{Astrophys. Space Sci} {\bfseries 165} (1990) 65}.

\bibitem{Sigl:1993ctk}
G.~Sigl and G.~Raffelt, \emph{{General kinetic description of relativistic mixed neutrinos}}, \href{https://doi.org/10.1016/0550-3213(93)90175-O}{\emph{Nucl. Phys. B} {\bfseries 406} (1993) 423}.

\bibitem{Sirera:1998ia}
M.~Sirera and A.~Perez, \emph{{Relativistic Wigner function approach to neutrino propagation in matter}}, \href{https://doi.org/10.1103/PhysRevD.59.125011}{\emph{Phys. Rev. D} {\bfseries 59} (1999) 125011} [\href{https://arxiv.org/abs/hep-ph/9810347}{{\ttfamily hep-ph/9810347}}].

\bibitem{Yamada:2000za}
S.~Yamada, \emph{{Boltzmann equations for neutrinos with flavor mixings}}, \href{https://doi.org/10.1103/PhysRevD.62.093026}{\emph{Phys. Rev. D} {\bfseries 62} (2000) 093026} [\href{https://arxiv.org/abs/astro-ph/0002502}{{\ttfamily astro-ph/0002502}}].

\bibitem{Vlasenko:2013fja}
A.~Vlasenko, G.M.~Fuller and V.~Cirigliano, \emph{{Neutrino Quantum Kinetics}}, \href{https://doi.org/10.1103/PhysRevD.89.105004}{\emph{Phys. Rev. D} {\bfseries 89} (2014) 105004} [\href{https://arxiv.org/abs/1309.2628}{{\ttfamily 1309.2628}}].

\bibitem{Volpe:2013uxl}
C.~Volpe, D.~V\"a\"an\"anen and C.~Espinoza, \emph{{Extended evolution equations for neutrino propagation in astrophysical and cosmological environments}}, \href{https://doi.org/10.1103/PhysRevD.87.113010}{\emph{Phys. Rev. D} {\bfseries 87} (2013) 113010} [\href{https://arxiv.org/abs/1302.2374}{{\ttfamily 1302.2374}}].

\bibitem{Serreau:2014cfa}
J.~Serreau and C.~Volpe, \emph{{Neutrino-antineutrino correlations in dense anisotropic media}}, \href{https://doi.org/10.1103/PhysRevD.90.125040}{\emph{Phys. Rev. D} {\bfseries 90} (2014) 125040} [\href{https://arxiv.org/abs/1409.3591}{{\ttfamily 1409.3591}}].

\bibitem{Kartavtsev:2015eva}
A.~Kartavtsev, G.~Raffelt and H.~Vogel, \emph{{Neutrino propagation in media: Flavor, helicity, and pair correlations}}, \href{https://doi.org/10.1103/PhysRevD.91.125020}{\emph{Phys. Rev. D} {\bfseries 91} (2015) 125020} [\href{https://arxiv.org/abs/1504.03230}{{\ttfamily 1504.03230}}].

\bibitem{Fiorillo:2024fnl}
D.F.G.~Fiorillo, G.G.~Raffelt and G.~Sigl, \emph{{Inhomogeneous Kinetic Equation for Mixed Neutrinos: Tracing the Missing Energy}}, \href{https://doi.org/10.1103/PhysRevLett.133.021002}{\emph{Phys. Rev. Lett.} {\bfseries 133} (2024) 021002} [\href{https://arxiv.org/abs/2401.05278}{{\ttfamily 2401.05278}}].

\bibitem{Fiorillo:2024wej}
D.F.G.~Fiorillo, G.G.~Raffelt and G.~Sigl, \emph{{Collective neutrino-antineutrino oscillations in dense neutrino environments?}}, \href{https://doi.org/10.1103/PhysRevD.109.043031}{\emph{Phys. Rev. D} {\bfseries 109} (2024) 043031} [\href{https://arxiv.org/abs/2401.02478}{{\ttfamily 2401.02478}}].

\bibitem{Mikheyev:1985zog}
S.P.~Mikheyev and A.{\relax{Yu}}.~Smirnov, \emph{{Resonance Amplification of Oscillations in Matter and Spectroscopy of Solar Neutrinos}}, {\emph{Sov. J. Nucl. Phys.} {\bfseries 42} (1985) 913}. {[{\em Yad. Fiz.} {\bf 42} (1985) 1441]}.

\bibitem{Wolfenstein:1977ue}
L.~Wolfenstein, \emph{{Neutrino oscillations in matter}}, \href{https://doi.org/10.1103/PhysRevD.17.2369}{\emph{Phys. Rev. D} {\bfseries 17} (1978) 2369}.

\bibitem{Bhattacharyya:2020jpj}
S.~Bhattacharyya and B.~Dasgupta, \emph{{Fast Flavor Depolarization of Supernova Neutrinos}}, \href{https://doi.org/10.1103/PhysRevLett.126.061302}{\emph{Phys. Rev. Lett.} {\bfseries 126} (2021) 061302} [\href{https://arxiv.org/abs/2009.03337}{{\ttfamily 2009.03337}}].

\bibitem{Zaizen:2022cik}
M.~Zaizen and H.~Nagakura, \emph{{Simple method for determining asymptotic states of fast neutrino-flavor conversion}}, \href{https://doi.org/10.1103/PhysRevD.107.103022}{\emph{Phys. Rev. D} {\bfseries 107} (2023) 103022} [\href{https://arxiv.org/abs/2211.09343}{{\ttfamily 2211.09343}}].

\bibitem{Ehring:2023lcd}
J.~Ehring, S.~Abbar, H.-T.~Janka, G.~Raffelt and I.~Tamborra, \emph{{Fast neutrino flavor conversion in core-collapse supernovae: A parametric study in 1D models}}, \href{https://doi.org/10.1103/PhysRevD.107.103034}{\emph{Phys. Rev. D} {\bfseries 107} (2023) 103034} [\href{https://arxiv.org/abs/2301.11938}{{\ttfamily 2301.11938}}].

\bibitem{Ehring:2023abs}
J.~Ehring, S.~Abbar, H.-T.~Janka, G.~Raffelt and I.~Tamborra, \emph{{Fast Neutrino Flavor Conversions Can Help and Hinder Neutrino-Driven Explosions}}, \href{https://doi.org/10.1103/PhysRevLett.131.061401}{\emph{Phys. Rev. Lett.} {\bfseries 131} (2023) 061401} [\href{https://arxiv.org/abs/2305.11207}{{\ttfamily 2305.11207}}].

\bibitem{Nagakura:2023jfi}
H.~Nagakura, L.~Johns and M.~Zaizen, \emph{{Bhatnagar-Gross-Krook subgrid model for neutrino quantum kinetics}}, \href{https://doi.org/10.1103/PhysRevD.109.083013}{\emph{Phys. Rev. D} {\bfseries 109} (2024) 083013} [\href{https://arxiv.org/abs/2312.16285}{{\ttfamily 2312.16285}}].

\bibitem{Xiong:2023vcm}
Z.~Xiong, M.-R.~Wu, S.~Abbar, S.~Bhattacharyya, M.~George and C.-Y.~Lin, \emph{{Evaluating approximate asymptotic distributions for fast neutrino flavor conversions in a periodic 1D box}}, \href{https://doi.org/10.1103/PhysRevD.108.063003}{\emph{Phys. Rev. D} {\bfseries 108} (2023) 063003} [\href{https://arxiv.org/abs/2307.11129}{{\ttfamily 2307.11129}}].

\bibitem{Cornelius:2023eop}
M.~Cornelius, S.~Shalgar and I.~Tamborra, \emph{{Perturbing fast neutrino flavor conversion}}, \href{https://doi.org/10.1088/1475-7516/2024/02/038}{\emph{JCAP} {\bfseries 02} (2024) 038} [\href{https://arxiv.org/abs/2312.03839}{{\ttfamily 2312.03839}}].

\bibitem{Johns:2024dbe}
L.~Johns, \emph{{Subgrid modeling of neutrino oscillations in astrophysics}},  \href{https://arxiv.org/abs/2401.15247}{{\ttfamily 2401.15247}}.

\bibitem{Abbar:2024ynh}
S.~Abbar, M.-R.~Wu and Z.~Xiong, \emph{{Application of neural networks for the reconstruction of supernova neutrino energy spectra following fast neutrino flavor conversions}}, \href{https://doi.org/10.1103/PhysRevD.109.083019}{\emph{Phys. Rev. D} {\bfseries 109} (2024) 083019} [\href{https://arxiv.org/abs/2401.17424}{{\ttfamily 2401.17424}}].

\bibitem{Fiorillo:2024qbl}
D.F.G.~Fiorillo and G.~Raffelt, \emph{{Fast flavor conversions at the edge of instability}},  \href{https://arxiv.org/abs/2403.12189}{{\ttfamily 2403.12189}}.

\bibitem{Xiong:2024pue}
Z.~Xiong, M.-R.~Wu, M.~George and C.-Y.~Lin, \emph{{Robust integration of fast flavor conversions in classical neutrino transport}},  \href{https://arxiv.org/abs/2403.17269}{{\ttfamily 2403.17269}}.

\bibitem{Fiorillo:2023mze}
D.F.G.~Fiorillo and G.G.~Raffelt, \emph{{Slow and fast collective neutrino oscillations: Invariants and reciprocity}}, \href{https://doi.org/10.1103/PhysRevD.107.043024}{\emph{Phys. Rev. D} {\bfseries 107} (2023) 043024} [\href{https://arxiv.org/abs/2301.09650}{{\ttfamily 2301.09650}}].

\bibitem{Johns:2023jjt}
L.~Johns, \emph{{Thermodynamics of oscillating neutrinos}},  \href{https://arxiv.org/abs/2306.14982}{{\ttfamily 2306.14982}}.

\bibitem{Johns:2024bob}
L.~Johns, \emph{{Ergodicity demystifies fast neutrino flavor instability}},  \href{https://arxiv.org/abs/2402.08896}{{\ttfamily 2402.08896}}.

\bibitem{Morinaga:2021vmc}
T.~Morinaga, \emph{{Fast neutrino flavor instability and neutrino flavor lepton number crossings}}, \href{https://doi.org/10.1103/PhysRevD.105.L101301}{\emph{Phys. Rev. D} {\bfseries 105} (2022) L101301} [\href{https://arxiv.org/abs/2103.15267}{{\ttfamily 2103.15267}}].

\bibitem{Dasgupta:2021gfs}
B.~Dasgupta, \emph{{Collective Neutrino Flavor Instability Requires a Crossing}}, \href{https://doi.org/10.1103/PhysRevLett.128.081102}{\emph{Phys. Rev. Lett.} {\bfseries 128} (2022) 081102} [\href{https://arxiv.org/abs/2110.00192}{{\ttfamily 2110.00192}}].

\bibitem{Pehlivan:2011hp}
Y.~Pehlivan, A.B.~Balantekin, T.~Kajino and T.~Yoshida, \emph{{Invariants of collective neutrino oscillations}}, \href{https://doi.org/10.1103/PhysRevD.84.065008}{\emph{Phys. Rev. D} {\bfseries 84} (2011) 065008} [\href{https://arxiv.org/abs/1105.1182}{{\ttfamily 1105.1182}}].

\bibitem{Johns:2019izj}
L.~Johns, H.~Nagakura, G.M.~Fuller and A.~Burrows, \emph{{Neutrino oscillations in supernovae: angular moments and fast instabilities}}, \href{https://doi.org/10.1103/PhysRevD.101.043009}{\emph{Phys. Rev. D} {\bfseries 101} (2020) 043009} [\href{https://arxiv.org/abs/1910.05682}{{\ttfamily 1910.05682}}].

\bibitem{Fiorillo:2023hlk}
D.F.G.~Fiorillo and G.G.~Raffelt, \emph{{Flavor solitons in dense neutrino gases}}, \href{https://doi.org/10.1103/PhysRevD.107.123024}{\emph{Phys. Rev. D} {\bfseries 107} (2023) 123024} [\href{https://arxiv.org/abs/2303.12143}{{\ttfamily 2303.12143}}].

\bibitem{Vlasov:1945}
A.A.~Vlasov, \emph{On the kinetic theory of an assembly of particles with collective interaction}, {\emph{J. Phys. USSR} {\bfseries 9} (1945) 25}.

\bibitem{VanKampen:1955}
N.G.~{Van Kampen}, \emph{{On the theory of stationary waves in plasmas}}, \href{https://doi.org/10.1016/S0031-8914(55)93068-8}{\emph{Physica} {\bfseries 21} (1955) 949}.

\bibitem{Case:1959}
K.M.~{Case}, \emph{{Plasma oscillations}}, \href{https://doi.org/10.1016/0003-4916(59)90029-6}{\emph{Annals of Physics} {\bfseries 7} (1959) 349}.

\bibitem{Landau:1946jc}
L.D.~Landau, \emph{{On the vibrations of the electronic plasma}}, {\emph{J. Phys. (USSR)} {\bfseries 10} (1946) 25}.

\bibitem{Nagakura:2022kic}
H.~Nagakura and M.~Zaizen, \emph{{Time-Dependent and Quasisteady Features of Fast Neutrino-Flavor Conversion}}, \href{https://doi.org/10.1103/PhysRevLett.129.261101}{\emph{Phys. Rev. Lett.} {\bfseries 129} (2022) 261101} [\href{https://arxiv.org/abs/2206.04097}{{\ttfamily 2206.04097}}].

\bibitem{Izaguirre:2016gsx}
I.~Izaguirre, G.~Raffelt and I.~Tamborra, \emph{{Fast Pairwise Conversion of Supernova Neutrinos: A Dispersion-Relation Approach}}, \href{https://doi.org/10.1103/PhysRevLett.118.021101}{\emph{Phys. Rev. Lett.} {\bfseries 118} (2017) 021101} [\href{https://arxiv.org/abs/1610.01612}{{\ttfamily 1610.01612}}].

\bibitem{Capozzi:2017gqd}
F.~Capozzi, B.~Dasgupta, E.~Lisi, A.~Marrone and A.~Mirizzi, \emph{{Fast flavor conversions of supernova neutrinos: Classifying instabilities via dispersion relations}}, \href{https://doi.org/10.1103/PhysRevD.96.043016}{\emph{Phys. Rev. D} {\bfseries 96} (2017) 043016} [\href{https://arxiv.org/abs/1706.03360}{{\ttfamily 1706.03360}}].

\bibitem{Capozzi:2019lso}
F.~Capozzi, G.~Raffelt and T.~Stirner, \emph{{Fast Neutrino Flavor Conversion: Collective Motion vs. Decoherence}}, \href{https://doi.org/10.1088/1475-7516/2019/09/002}{\emph{JCAP} {\bfseries 09} (2019) 002} [\href{https://arxiv.org/abs/1906.08794}{{\ttfamily 1906.08794}}].

\bibitem{VanKampen:1955wh}
N.G.~Van~Kampen, \emph{{On the theory of stationary waves in plasmas}}, \href{https://doi.org/10.1016/S0031-8914(55)93068-8}{\emph{Physica} {\bfseries 21} (1955) 949}.

\bibitem{case1959plasma}
K.M.~Case, \emph{Plasma oscillations}, {\emph{Annals of Physics} {\bfseries 7} (1959) 349}.

\bibitem{Sagan:1993es}
D.~Sagan, \emph{{On the physics of Landau damping}}, \href{https://doi.org/10.1119/1.17547}{\emph{Am. J. Phys.} {\bfseries 62} (1994) 450}.

\bibitem{pomeranchuk1959stability}
I.{\relax{Ia}}.~Pomeranchuk, \emph{{On the stability of a Fermi liquid}}, {\emph{Sov. Phys. JETP} {\bfseries 8} (1959) 361}. [Translated from {\em J. Exptl. Theoret. Phys. (U.S.S.R.)} {\bf 3} (1958) 524--525], \href{http://www.jetp.ras.ru/cgi-bin/dn/e_008_02_0361.pdf}{Link}.

\bibitem{Airen:2018nvp}
S.~Airen, F.~Capozzi, S.~Chakraborty, B.~Dasgupta, G.~Raffelt and T.~Stirner, \emph{{Normal-mode Analysis for Collective Neutrino Oscillations}}, \href{https://doi.org/10.1088/1475-7516/2018/12/019}{\emph{JCAP} {\bfseries 12} (2018) 019} [\href{https://arxiv.org/abs/1809.09137}{{\ttfamily 1809.09137}}].

\bibitem{lifshitz2013statistical}
E.M.~Lifshitz and L.P.~Pitaevskii, \emph{Statistical physics: theory of the condensed state}, vol.~9, Elsevier (2013).

\bibitem{kramers1928diffusion}
H.A.~Kramers, \emph{La diffusion de la lumi{\`e}re par les atomes}, {\emph{Atti del Congresso Internationale dei Fisici (Como)} {\bfseries 2} (1927) 545}.

\bibitem{deL.Kronig:26}
R.~de~L.~Kronig, \emph{{On the Theory of Dispersion of X-Rays}}, \href{https://doi.org/10.1364/JOSA.12.000547}{\emph{J. Opt. Soc. Am.} {\bfseries 12} (1926) 547}.

\bibitem{Kirzhnits:1989zz}
D.A.~Kirzhnits, \emph{{General properties of electromagnetic response functions}}, \href{https://doi.org/10.1016/B978-0-444-87366-8.50008-4}{\emph{Mod. Probl. Condens. Matter Sci.} {\bfseries 24} (1989) 41}.

\bibitem{Gamow:1928zz}
G.~Gamow, \emph{{Zur Quantentheorie des Atomkernes}}, \href{https://doi.org/10.1007/BF01343196}{\emph{Z. Phys.} {\bfseries 51} (1928) 204}.

\bibitem{Fiorillo:2024}
D.F.G.~Fiorillo and G.G.~Raffelt, \emph{{Theory of neutrino fast flavor evolution. II. Solutions at the edge of instability.}},  2024.
\newblock Work in Progress.

\bibitem{Landau:1975pou}
L.D.~Landau and E.M.~Lifschits, \emph{{The Classical Theory of Fields}}, vol.~Volume 2 of \emph{Course of Theoretical Physics}, Pergamon Press, Oxford (1975).

\bibitem{thorne2017modern}
K.S.~Thorne and R.D.~Blandford, \emph{Modern classical physics: optics, fluids, plasmas, elasticity, relativity, and statistical physics}, Princeton University Press (2017).

\bibitem{Penrose:1960}
O.~{Penrose}, \emph{{Electrostatic Instabilities of a Uniform Non-Maxwellian Plasma}}, \href{https://doi.org/10.1063/1.1706024}{\emph{Physics of Fluids} {\bfseries 3} (1960) 258}.

\bibitem{pines1952collective}
D.~Pines and D.~Bohm, \emph{{A Collective Description of Electron Interactions: II.~Collective vs Individual Particle Aspects of the Interactions}}, \href{https://doi.org/10.1103/PhysRev.85.338}{\emph{Phys. Rev.} {\bfseries 85} (1952) 338}.

\bibitem{bohm1949theory}
D.~Bohm and E.P.~Gross, \emph{{Theory of Plasma Oscillations. A. Origin of Medium-Like Behavior}}, \href{https://doi.org/10.1103/PhysRev.75.1851}{\emph{Phys. Rev.} {\bfseries 75} (1949) 1851}.

\bibitem{ginzburg1970propagation}
V.~Ginzburg, \emph{The Propagation of Electromagnetic Waves in Plasmas}, Commonwealth and International Library, Pergamon Press (1970).

\bibitem{shafranov1958propagation}
V.D.~Shafranov, \emph{Propagation of an electromagnetic field in a medium with spatial dispersion}, {\emph{Sov. Phys. JETP} {\bfseries 7} (1958) 1019}. [Translated from {\em J. Exptl. Theoret. Phys. (U.S.S.R.)} {\bf 34} (1958) 1475--1489], \href{http://www.jetp.ras.ru/cgi-bin/dn/e_007_06_1019.pdf}{Link}.

\bibitem{Altherr:1992mf}
T.~Altherr and U.~Kraemmer, \emph{{Gauge field theory methods for ultradegenerate and ultrarelativistic plasmas}}, \href{https://doi.org/10.1016/0927-6505(92)90014-Q}{\emph{Astropart. Phys.} {\bfseries 1} (1992) 133}.

\bibitem{Braaten:1993jw}
E.~Braaten and D.~Segel, \emph{{Neutrino energy loss from the plasma process at all temperatures and densities}}, \href{https://doi.org/10.1103/PhysRevD.48.1478}{\emph{Phys. Rev. D} {\bfseries 48} (1993) 1478} [\href{https://arxiv.org/abs/hep-ph/9302213}{{\ttfamily hep-ph/9302213}}].

\bibitem{Dasgupta:2009mg}
B.~Dasgupta, A.~Dighe, G.G.~Raffelt and A.Y.~Smirnov, \emph{{Multiple Spectral Splits of Supernova Neutrinos}}, \href{https://doi.org/10.1103/PhysRevLett.103.051105}{\emph{Phys. Rev. Lett.} {\bfseries 103} (2009) 051105} [\href{https://arxiv.org/abs/0904.3542}{{\ttfamily 0904.3542}}].

\bibitem{Shalgar:2020xns}
S.~Shalgar and I.~Tamborra, \emph{{Dispelling a myth on dense neutrino media: fast pairwise conversions depend on energy}}, \href{https://doi.org/10.1088/1475-7516/2021/01/014}{\emph{JCAP} {\bfseries 01} (2021) 014} [\href{https://arxiv.org/abs/2007.07926}{{\ttfamily 2007.07926}}].

\bibitem{Johns:2021qby}
L.~Johns, \emph{{Collisional Flavor Instabilities of Supernova Neutrinos}}, \href{https://doi.org/10.1103/PhysRevLett.130.191001}{\emph{Phys. Rev. Lett.} {\bfseries 130} (2023) 191001} [\href{https://arxiv.org/abs/2104.11369}{{\ttfamily 2104.11369}}].

\bibitem{Xiong:2022zqz}
Z.~Xiong, L.~Johns, M.-R.~Wu and H.~Duan, \emph{{Collisional flavor instability in dense neutrino gases}}, \href{https://doi.org/10.1103/PhysRevD.108.083002}{\emph{Phys. Rev. D} {\bfseries 108} (2023) 083002} [\href{https://arxiv.org/abs/2212.03750}{{\ttfamily 2212.03750}}].

\bibitem{Liu:2023pjw}
J.~Liu, M.~Zaizen and S.~Yamada, \emph{{Systematic study of the resonancelike structure in the collisional flavor instability of neutrinos}}, \href{https://doi.org/10.1103/PhysRevD.107.123011}{\emph{Phys. Rev. D} {\bfseries 107} (2023) 123011} [\href{https://arxiv.org/abs/2302.06263}{{\ttfamily 2302.06263}}].

\bibitem{Lin:2022dek}
Y.-C.~Lin and H.~Duan, \emph{{Collision-induced flavor instability in dense neutrino gases with energy-dependent scattering}}, \href{https://doi.org/10.1103/PhysRevD.107.083034}{\emph{Phys. Rev. D} {\bfseries 107} (2023) 083034} [\href{https://arxiv.org/abs/2210.09218}{{\ttfamily 2210.09218}}].

\bibitem{Johns:2022yqy}
L.~Johns and Z.~Xiong, \emph{{Collisional instabilities of neutrinos and their interplay with fast flavor conversion in compact objects}}, \href{https://doi.org/10.1103/PhysRevD.106.103029}{\emph{Phys. Rev. D} {\bfseries 106} (2022) 103029} [\href{https://arxiv.org/abs/2208.11059}{{\ttfamily 2208.11059}}].

\bibitem{Padilla-Gay:2022wck}
I.~Padilla-Gay, I.~Tamborra and G.G.~Raffelt, \emph{{Neutrino fast flavor pendulum. II. Collisional damping}}, \href{https://doi.org/10.1103/PhysRevD.106.103031}{\emph{Phys. Rev. D} {\bfseries 106} (2022) 103031} [\href{https://arxiv.org/abs/2209.11235}{{\ttfamily 2209.11235}}].

\bibitem{Fiorillo:2023ajs}
D.F.G.~Fiorillo, I.~Padilla-Gay and G.G.~Raffelt, \emph{{Collisions and collective flavor conversion: Integrating out the fast dynamics}}, \href{https://doi.org/10.1103/PhysRevD.109.063021}{\emph{Phys. Rev. D} {\bfseries 109} (2024) 063021} [\href{https://arxiv.org/abs/2312.07612}{{\ttfamily 2312.07612}}].

\end{thebibliography}\endgroup

\end{document}